\RequirePackage[normalem]{ulem} %DIF PREAMBLE
\RequirePackage{color}\definecolor{RED}{rgb}{1,0,0}\definecolor{BLUE}{rgb}{0,0,1} %DIF PREAMBLE\providecommand{\DIFadd}[1]{{\protect\color{blue}\uwave{#1}}} %DIF PREAMBLE
\documentclass[12pt,usenatbib]{mnras}
\usepackage{url,ulem,times,graphicx,amsmath,amsfonts,amssymb,color,epsfig,epstopdf}

\usepackage{graphics}
\usepackage{epsf}
\usepackage{bm}
\usepackage{color}
\usepackage{rotating}
\usepackage[T1]{fontenc}
\usepackage{ae,aecompl}
 \usepackage{array,multirow}
\usepackage[percent]{overpic}
\usepackage{tikz}
\def\mnras{MNRAS}

\def\physrep{Physics Reports}

\def\araa{Annual Review of Astronomy and Astrophysics}
\def\aap{A\&A}

\def\apj{ApJ}
\def\apjs{ApJS}
\def\apjl{ApJL}
\def\aj{AJ}

\def\lsim{\mathrel{\lower0.6ex\hbox{$\buildrel {\textstyle <}
 \over {\scriptstyle \sim}$}}}
\def\gsim{\mathrel{\lower0.6ex\hbox{$\buildrel {\textstyle >}
 \over {\scriptstyle \sim}$}}}

\begin{document}

\title[The Hestia Simulations ]{The {\sc Hestia} project:  simulations of the Local Group}
\author[Libeskind et al]
{\newauthor Noam I. Libeskind$^{1,2}$ Edoardo Carlesi$^{1}$, Rob J. J. Grand$^3$,  Arman Khalatyan$^1$,  \newauthor Alexander Knebe$^{4,5,6}$, Ruediger Pakmor$^3$, Sergey Pilipenko$^7$, Marcel S. Pawlowski$^1$, \newauthor Martin Sparre$^{1,8}$, Elmo Tempel$^9$,  Peng Wang$^1$, H\'el\`ene M. Courtois$^{2}$, Stefan Gottl\"{o}ber$^{1}$, \newauthor Yehuda Hoffman$^{10}$, Ivan Minchev$^1$, Christoph Pfrommer$^1$, Jenny G. Sorce$^{11,12,1}$, \newauthor Volker Springel$^3$, Matthias Steinmetz$^{1}$, R. Brent Tully$^{13}$, Mark Vogelsberger$^{14}$,  \newauthor Gustavo Yepes$^{4,5}$ \\
$^1$Leibniz-Institut f\"ur Astrophysik Potsdam (AIP), An der Sternwarte 16, D-14482 Potsdam, Germany\\
$^2$University of Lyon, UCB Lyon 1, CNRS/IN2P3, IUF, IP2I Lyon, France\\
 $^3$Max-Planck-Institut f\"ur Astrophysik, Karl-Schwarzschild-Str. 1, D-85748, Garching, Germany\\
 $^4$Departamento de F\'isica Te\'{o}rica, M\'{o}dulo 15, Facultad de Ciencias, Universidad Aut\'{o}noma de Madrid, 28049 Madrid, Spain\\
 $^5$Centro de Investigaci\'{o}n Avanzada en F\'isica Fundamental (CIAFF), Facultad de Ciencias, Universidad Aut\'{o}noma de Madrid, \\28049 Madrid, Spain\\
$^{6}$International Centre for Radio Astronomy Research, University of Western Australia, 35 Stirling Highway, Crawley,\\ Western Australia 6009, Australia\\
$^7$P.N. Lebedev Physical Institute of Russian Academy of Sciences, 84/32 Profsojuznaja Street, 117997 Moscow, Russia \\
$^8$Potsdam University\\
$^9$Tartu Observatory, University of Tartu, Observatooriumi 1, 61602 T\~oravere, Estonia\\
$^{10}$Racah Institute of Physics, Hebrew University, Jerusalem 91904, Israel\\
$^{11}$Univ. Lyon, ENS de Lyon, Univ. Lyon I, CNRS, Centre de Recherche Astrophysique de Lyon, UMR5574, F-69007, Lyon, France\\
$^{12}$Univ Lyon, Univ Lyon-1, Ens de Lyon, CNRS, Centre de Recherche Astrophysique de Lyon
UMR5574, F-69230, Saint-Genis-Laval, France\\
 $^{13}$Institute for Astronomy (IFA), University of Hawaii, 2680 Woodlawn Drive, HI 96822, USA\\
 $^{14}$Department of Physics, Massachusetts Institute of Technology, Cambridge, MA 02139, USA
   }
\date{Accepted --- . Received ---; in original form ---}
\pagerange{\pageref{firstpage}--\pageref{lastpage}} \pubyear{2020}
\maketitle

 \begin{abstract}
We present the {\sc Hestia} simulation suite: High-resolutions Environmental Simulations of The Immediate Area, a set of cosmological simulations of the Local Group. Initial conditions constrained by the observed peculiar velocity of nearby galaxies are employed to accurately simulate the local cosmography. Halo pairs that resemble the Local Group are found in low resolutions constrained, dark matter only simulations, and selected for higher resolution magneto hydrodynamic simulation using the {\sc Arepo} code. Baryonic physics follows the {\sc Auriga} model of galaxy formation. The simulations contain a high resolution region of 3-5~Mpc in radius from the Local Group midpoint embedded in the correct cosmographic landscape. Within this region a simulated Local Group consisting of a Milky Way and Andromeda like galaxy forms, whose description is in excellent agreement with observations. The simulated Local Group galaxies resemble the Milky Way and Andromeda in terms of their halo mass, mass ratio, stellar disc mass, morphology separation, relative velocity, rotation curves, bulge-disc morphology, satellite galaxy stellar mass function, satellite radial distribution and in some cases, the presence of a Magellanic cloud like object. Because these simulations properly model  the Local Group in their cosmographic context, they provide a testing ground for questions where environment is thought to play an important role.

\end{abstract}

\section{Introduction}
\label{section:intro}
\label{firstpage}
Astronomers and astrophysicists have long used nearby objects as guides for understanding more distant ones. The Earth is often taken as a paradigmatic rocky planet, the Sun as a model star, and the Milky Way (MW) as an archetypal spiral galaxy. This pedagogy is natural given that the Earth, the Sun and the Galaxy are observed in much greater detail than their more distant counterparts. However, such an approach will always beg the questions: {\it how prototypical are these objects? How responsible is it to extrapolate what we learn here, to there?} Just because they are near and lend themselves naturally to the kind of observation one can only dream of obtaining elsewhere, needn't mean they are representative of their astronomical object ``class'' as a whole. Such a debate constitutes the essence of an observational science.

Furthermore, a number of observations of the MW have recently exacerbated claims that perhaps it is unusual. However (driven by the underlying axiom of our field) cosmologists are loathe to invoke any anti-copernican explanation for confusing observations. Specifically,  many of the satellites of the MW -- small dwarf galaxies many with low surface brightness that orbit inside its dark matter halo, have posed a number of questions that remain either fully or partially unresolved. For example, the ``classical'' dwarf galaxies are distributed in a highly anisotropic ``Great Pancake'' \citep{2005A&A...431..517K,2005MNRAS.363..146L} or ``disc of Satellites'' \citep[see][for a review]{2018MPLA...3330004P}. Another satellite issue was highlighted by \cite{2017ApJ...847....4G} who found that the MW's satellite population, when compared with observational ``analogues'' (based on K-band luminosity), have significantly lower star formation rates than expected. The so-called ``Missing satellite'' problem \citep{1999ApJ...524L..19M,1999ApJ...522...82K} is perhaps the most (over) studied of these and refers to the over-prediction of dark matter satellites in the $\Lambda$CDM model of structure formation, when compared with the bright observed satellites of the MW (for a non exhaustive list of possible remedies see \citealt{1992MNRAS.256P..43E,2002MNRAS.333..156B,2002MNRAS.329..813K,2012MNRAS.420.2318L,2013ApJ...765...22B,2016MNRAS.456...85S,2016MNRAS.457.1931S}). Similarly, \cite{Boylan11} showed that the internal densities of the most massive simulated substructures around MW mass haloes is too large that they can not have been rendered ``dark'' via reionization or photo evaporation processes, but these are not observed. It is clear that in the dwarf galaxies regime, objects that are best observed in the Local Group (LG), a number of discrepancies with theoretical predictions have been found.

The problem may be that the starting point for such studies is to assume the MW is a proto-typical galaxy that lives in a $\sim10^{12}M_{\odot}$ halo. This may or may not be the case. What is known is that galaxy evolution is regulated by environment either via merger rates or the availability of a gas reservoir from which stars may form. \cite{1980ApJ...236..351D} first pointed out that redder early-type galaxies consisting of older stellar populations tend to reside in more dense regions than bluer late type discs. Galaxies in voids are about 30\% more metal poor than the general population \citep{2011AstBu..66..255P}. It is also empirically evidenced that neighbouring galaxies have correlated specific star formation rates \citep{2006MNRAS.366....2W,2010MNRAS.409..491K,2013MNRAS.430.1447K,2015ApJ...800...24K}, a phenomenon termed {\it galactic conformity} \citep[see also][]{2016MNRAS.461.2135H}. Galaxy spin and angular momentum is also well known to be shaped by environment  \citep[see][among others]{2004ApJ...613L..41N,2013ApJ...775L..42T,2016MNRAS.457..695P,2017MNRAS.468L.123W,2018MNRAS.474.5437L,2019MNRAS.487.1607G}. The so called ``nature vs. nurture'' debate can be summarised as: how does environment regulate galaxy formation and evolution?

In order to address these questions in the context of the LG one must, of course, simulate galaxy systems that resemble the LG. The most common (albeit slightly simplistic) approach is to simulate pairs of galaxies. Studies such as \cite{2014MNRAS.438.2578G}, \cite{2016MNRAS.457.1931S} and \cite{2020MNRAS.493.2596F}, among others have  investigated pairs of galaxies that resemble the LG geometrically and physically. Often the results suggest that the nature of the LG itself is not enough to account for differences between isolated MW like haloes and paired ones.

 But where these studies fall short is that the environment -- the cosmographic landscape --  is not accounted for. Namely such studies produce LG like pairs in uncontrolled, random environments. However our cosmographic landscape - a supergalactic plane, the existence of the Virgo Cluster, a filament funnelling material towards it, the evacuation from the local void, all influence the formation of the LG.  \cite{2015MNRAS.452.1052L,2019MNRAS.490.3786L} suggested that the planes of satellites observed in the LG are aligned with the reconstructed large-scale structure, hinting at a causal relationship. Similarly \cite{2013MNRAS.436.2096S} pointed towards the relationship between planes of satellites and evacuation from the Local Void \citep[see][]{2018NatAs...2..680H,2019ApJ...880...24T}. \cite{2014MNRAS.440..405M} found evidence, via a ``council of giant'' galaxies, that the LG is dynamically connected to its local environment \citep[see also][]{2020MNRAS.494.2600N} that stretches well into the quasi-linear regime (3-5~Mpc) and is influenced by linear scales as well  \citep[i.e. the Virgo cluster, see][]{2008MNRAS.389.1001L}. \cite{2011MNRAS.411.1525L} showed that satellites are accreted from the direction pointing to the Virgo cluster in simulations, later confirmed more generically in the context of tidal forces \cite[][see also \citealt{2017MNRAS.472.4099K}]{2014MNRAS.443.1274L,2015ApJ...813....6K,2018MNRAS.473.1562W}. These studies (and more) have emphasised the importance for the local cosmography on the formation of the LG, and the MW.

A commitment to modelling the environment in studies of the LG was made by the {\sc clues} collaboration\footnote{\url{www.clues-project.org}} \citep[][]{2010arXiv1005.2687G,2010MNRAS.401.1889L,2016MNRAS.458..900C,2016MNRAS.455.2078S} which recognised this often overlooked  aspect of most theoretical studies of the LGs. Simulations emanating from the {\sc Clues} collaboration have already demonstrated the importance of environment in studies which, for example, focused on  the preferred directions of accretion \citep{2011MNRAS.411.1525L}, the trajectory of backsplash galaxies \citep{2011MNRAS.412..529K}, LG mass estimators \citep{DiCintio12b}, cosmic web stripping of LG dwarfs \citep{2013ApJ...763L..41B},  and the reionization of the LG \citep{2013ApJ...777...51O,2014ApJ...794...20O,2018MNRAS.477..867D}.  The purpose of this paper is to combine a contemporary magento-hydrodynamic (MHD) treatment of cosmic gas with the state-of-the-art Auriga galaxy formation model \citep{2017MNRAS.467..179G} run on initial conditions (ICs) constrained by cosmographic observations thus destined to produce an object that resembles the LG in two important ways: internally (MW and M31 type galaxies, at the correct separation, etc.) and externally (the correct large scale structure).

The simulations presented here thus constitute an extension and improvement on a canon of previous work. A number of improvements have been made regarding  the constrained simulation algorithm, the MHD galaxy formation models, as well as the input data. These are the first hydrodynamic zoom-in, $z=0$ simulations run on ICs constrained by peculiar velocities presented in the CosmicFlows-2 catalogue (\citealt{2013AJ....146...86T}, see Section.~\ref{sec:wf}). The CosmicFlows database and catalogue have been used extensively before to study the local universe, including the cold spot repeller \citep{2017ApJ...847L...6C}, the Vela super cluster \citep{2019MNRAS.490L..57C}, the bulk flow \citep{2015MNRAS.449.4494H}, and the Virgo cluster \citep{2016MNRAS.460.2015S,2019MNRAS.486.3951S,2018A&A...614A.102O}. As will be demonstrated below, producing such constrained simulations is not trivial but will provide an optimum testing ground for understanding both the role of our environment in shaping the LG as well as the holistic formation of LG galaxies. 

\section{Modelling and simulation method}
This section includes background material on the constrained simulation method, the MHD and the galaxy formation algorithm.
\subsection{Wiener Filter reconstructions, CosmicFlows and IC generation}
\label{sec:wf}

In this section the technique used to generate ICs that are on one hand consistent with the $\Lambda$CDM prior, and on the other constrained by observational data to reproduce the bulk cosmographic landscape of the local Universe, are explained. For a comprehensive pedagogical explanation of the technique tersely summarized and augmented below, the reader is referred to  \cite{2009LNP...665..565H}.

In the paradigm adopted here -- the $\Lambda$CDM model -- structures form via
gravitational instability.  At high redshifts the (over)density and peculiar
velocity fields constitute Gaussian random fields of primordial fluctuations
\citep[i.e.][]{1970A&A.....5...84Z,1986ApJ...304...15B}. The emergence of
non-linear structure via gravitational instability drives the large-scale
structure away from Gaussianity, yet on scales down to a  few Mpc the velocity
field remains close to being a Gaussian random field. It follows that given a
measurement of the peculiar velocity field, the linear Bayesian tool of the
Wiener Filter (WF) and Constrained Realizations (CR)
\citep{1991ApJ...380L...5H,1995ApJ...449..446Z}  are optimal for the reconstruction of the large-scale structure and peculiar velocities \citep{1999ApJ...520..413Z}.

{\bf Sampling:} The process of generating constrained ICs commences with
sampling of the peculiar velocity field via direct measurements of galaxy
distances and redshifts. Peculiar velocities are then obtained by $v_{\rm pec}
=v_{\rm tot}  - H_{0}\times d $ where $d$ is the measured distance, $H_{0}$ is the Hubble expansion rate today and $v_{\rm tot}$ is the total velocity of the galaxy with respect to the CMB. In the following we use the publicly available CosmicFlows-2 (CF2) catalogue \citep{2013AJ....146...86T} which consists of over 8,000 direct distance measurements, out to a distance of $\sim$250~Mpc. The median distance of the catalogue is $\sim70$~Mpc.

{\bf Grouping:} The CF2 catalog includes a considerable number of galaxies
that are members of groups and clusters, the velocities of which are affected by
small-scale non-linear virial motions.  Grouping these galaxies - namely
collapsing data points identified as belonging to the same group or cluster into
a single data point, provides an effective remedy to the problem \citep{2017MNRAS.468.1812S} as it filters
the non-linear component of the large scale velocity field. Here we
have used the original grouping proposed for CF2 (Brent Tully, private
communication). This grouping  reduces the  size of the CF2 catalog to 
4837 data points.

{\bf Bias minimization:} It is well known that observational catalogs are inherently biased, often referred to as  ``Malmquist'' bias \citep[e.g. see section 6.4 of][]{1995PhR...261..271S}. One of the main issues dealt with here is that the observational error is on the distance modulus which, when converted in to a distance, translates in to a log-normal error distribution. This often leads to spurious infall in the absence of simulated galaxy clusters, if not properly corrected. Different types of biases have been discussed in \cite{2015MNRAS.450.2644S} where an 
iterative method to minimize the spurious infall on to the Local Volume, to reduce non-Gaussianities in the radial peculiar velocity distribution and to ensure the correct distribution of galaxy clusters \citep[including our closest cluster, Virgo. See ][]{2016MNRAS.460.2015S,2018MNRAS.478.5199S}, has been designed applied. The result is a new distribution of radial peculiar velocities, their uncertainties and corresponding distances. This distribution has been used in the next steps.

{\bf Displacement corrections:} The WF/CRs algorithm and the linear mapping to
the young universe are done within the Eulerian framework, in which
displacements are neglected. Yet, the growth of structure involves the
displacements of mass elements over cosmological time - in first approximation
the so-called Zeldovich displacements. The reversed Zeldovich approximation
has been devised to undo these displacements and is applied here to bring back
the data points close to their primordial positions with the 3D reconstructed cosmic
displacement field \citep{2013MNRAS.430..888D,2014MNRAS.437.3586S}.  

{\bf Constrained initial conditions:}  Density fields constrained by these modified observational peculiar velocities are constructed using the CR/WF technique \citep{1991ApJ...380L...5H,1999ApJ...520..413Z}. At this step the grouped and bias minimized peculiar velocity constraints are applied to a random (density and velocity) realization used to restore statistically the unconstrained structures. In fact, in the presence of the very noisy and incomplete velocity data the WF alone would provide no density fluctuations at all in most parts of the simulation volume (where there is no data the WF returns the mean field).  Finally, the density fields are rescaled  to build constrained ICs in the simulation box.

{\bf Small scale power:} Structures on scales smaller than that of
shell-crossing are predominantly random and are not directly affected by the imposed constraints. The desired reconstructed LG lies within this random domain and while its structure is not directly constrained by the CF2 data, the boundary conditions on the LG and the tidal field that affects it are strongly constrained by the data. The construction of  the ICs for the {\sc Hestia} LG is decomposed  into the construction of the long waves by the CR/WF algorithm described above and adding of short waves as random realizations assuming the same $\Lambda$CDM power spectrum \citep{2016MNRAS.458..900C}. Naturally, not all of these random realisation contain an object which could be identified as a LG. In Sect. 3.2 we describe the identification algorithm for the  ICs of the high resolution LG candidates.

\subsection{Simulation code}
\label{sec:hydro}
In this section we summarise the properties of the \textsc{Arepo} code we use and the \textsc{Auriga} galaxy formation model. A publicly available version of the {\sc arepo} code is available \citep{2019arXiv190904667W}\footnote{https://arepo-code.org/}.

\subsubsection{Magnetohydrodynamics with the moving-mesh code {\sc Arepo}}

The {\sc arepo} code \citep{2010MNRAS.401..791S,2016MNRAS.455.1134P} solves the ideal magnetohydrodynamics (MHD) equations on an unstructured Voronoi-mesh with a second order finite volume scheme. The Voronoi mesh is generated from a set of mesh-generating points, each of them creating one cell. The mesh-generating points move with the gas velocity, which reduces the mass flux over interfaces and the advection errors compared to static mesh codes, making the code quasi-Lagrangian. In addition, mesh-generating points of highly distorted cells are slowly moved towards their cell centers to make sure that the mesh stays regular \citep{2013MNRAS.436.3031V}. Additional explicit refinement and derefinement is used to control the properties of the cells where appropriate. By default, the mass of each cell is kept within a factor of two of a target mass resolution. However, {\sc arepo} in general allows for additional refinement criteria, e.g. to set a minimum cell size in the circumgalactic medium \citep{2019MNRAS.482L..85V}.

The fluxes over interfaces are computed in the moving frame of the interface using the HLLD Riemann solver \citep{2011MNRAS.418.1392P,2013MNRAS.432..176P}. The divergence of the magnetic field is controlled with the 8-wave scheme \citep{1999JCoPh.154..284P}. Time integration is done via a second order accurate scheme. {\sc arepo} allows a hierarchy of timesteps below the global timestep, such that every resolution element is integrated on the timestep required by its own local timestep criterion  \citep{2010MNRAS.401..791S}.

Self-gravity and other source terms are coupled to the ideal MHD equations by operator splitting. Gravitational forces are computed using a hybrid TreePM technique \citep{2005MNRAS.364.1105S} with two Fourier mesh levels, one for the full box and one centered on the high resolution region (see below).

\subsubsection{The Auriga galaxy formation model}
\label{sec:auriga}

The Auriga galaxy formation model \citep{2017MNRAS.467..179G}, which is based on the Illustris model \citep{2013MNRAS.436.3031V}, implements the most important physical processes relevant for the formation and evolution of galaxies. It includes cooling of gas via primordial and metal cooling \citep{2013MNRAS.436.3031V} and a spatially uniform UV background \citep{2013MNRAS.436.3031V}. The interstellar medium is described by a subgrid model for a two-phase medium in which cold star-forming clouds are embedded in a hot volume-filling medium \citep{2003MNRAS.339..289S}. Gas that is denser than $n_\mathrm{thres}=0.13 {\rm ~cm}^{-3}$ forms stars following a Schmidt-type star formation law. Star formation itself is done stochastically and creates star particles with the target gas mass that represent single stellar populations (SSP). The model includes mass loss and metal return from asymptotic giant branch (AGB) stars, core-collapse supernovae, and Type Ia supernovae that are distributed in the cells around a star particle.

Galactic winds are implemented by creating a wind with a given velocity and mass loading just outside the star-forming phase. At the technical level, this non-local momentum feedback is realised by creating temporary wind particles that are launched isotropically from sites of star formation, dumping their momentum, mass and energy into a local cell once they encounter gas with a density less than $0.05\,n_\mathrm{thres}$ \citep{2017MNRAS.467..179G}. Despite being launched isotropically, this model generates coherent bipolar outflows for disc galaxies at low redshift \citep{2019MNRAS.490.4786G,2019MNRAS.490.3234N}. The Auriga model also follows the formation and growth of supermassive black holes and includes their feedback as active galactic nuclei (AGN).

Magnetic fields are seeded as uniform seed fields at $z=127$ with a comoving field strength of  $10^{-14}$G. As the magnetic fields in galaxies are quickly amplified by an efficient turbulent dynamo at high redshift \citep{2014ApJ...783L..20P,2017MNRAS.469.3185P} they lose any memory of the initial seed field of the simulation. Note however, that the magnetic field between galaxies reflects the initial the seed field, and is therefore not a prediction of the simulations \citep{2015MNRAS.453.3999M}.

The Auriga model has been used successfully to study a large number of questions on the formation and evolution of Milky Way-like galaxies and their satellite galaxies. It reproduces many general properties of the stellar component of Milky Way-like galaxies  \citep{2017MNRAS.467..179G,2018MNRAS.474.3629G}, as well as their gas \citep{2017MNRAS.466.3859M}, their stellar halo \citep{2019MNRAS.485.2589M}, and their satellite systems \citep{2018MNRAS.478..548S,2020arXiv200212043F}. Moreover, the evolution of the magnetic fields in the disc and the circum-galactic medium has been studied extensively for the Auriga simulations and found to be in excellent agreement with observations of the Milky Way and other nearby disc galaxies \citep{2017MNRAS.469.3185P,2018MNRAS.481.4410P,2019arXiv191111163P}.

Finally, the Auriga model is very similar in the mass range of Milky Way-like galaxies to the IllustrisTNG model \citep{2017MNRAS.465.3291W,2018MNRAS.473.4077P} that has been found to quite successfully reproduce the properties of the full galaxy population in a representative part of the universe \citep[see, e.g.][]{2018MNRAS.475..676S}.

\section{Analysis}
All simulations in this work assume a cosmology consistent with the best fit \cite{2014A&A...571A..16P} values:  $\sigma_{8} = 0.83$ and $H_{0} = 100 h$ km s$^{-1}$ Mpc$^{-1}$ where $h=0.677$. We adopt $\Omega_{\Lambda} = 0.682$ throughout; for DM only runs $\Omega_{M} = 0.318$, while for hydrodynamic runs  $\Omega_{M} = 0.270$ and  $\Omega_{b} = 0.048$.

\begin{figure}
 \includegraphics[width=20pc]{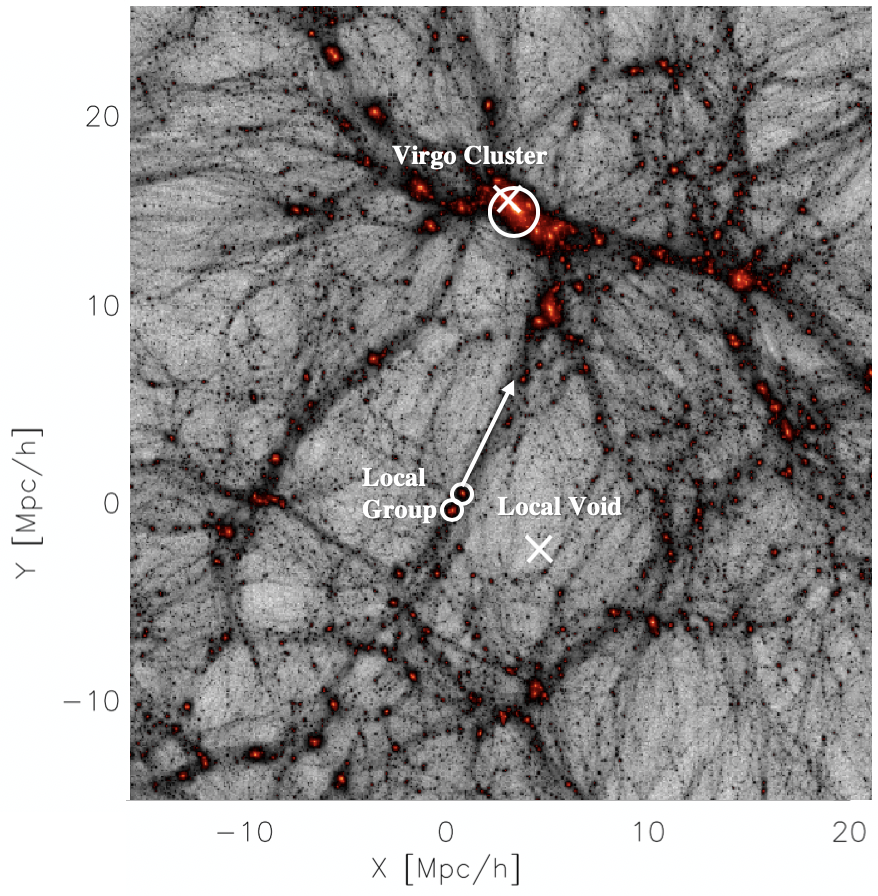}
\caption{A rendering of the density field in a typical constrained simulation. The position of the observed Virgo cluster and the center of the Local Void is demarcated by two white X's. The large white circle shows the position of the simulated Virgo cluster, while the region surrounding the Local Void is also clearly empty. A white arrow shows the direction of the local filament as determined by the velocity-shear tensor analysis of the reconstructed local Universe in \citet{2015MNRAS.452.1052L}. The filament direction is not perfectly aligned with the direction towards the Virgo cluster. The observed and simulated Local Group both reside at the super galactic origin, denoted by the two small white circles.}
 \label{fig:CSbox}
 \end{figure}

\subsection{Post-processing of the simulation output}
\label{sec:pp}
In this section the post-processing halo finder (and accompanying tools) that are used throughout this work to identify haloes, subhaloes, galaxies and merger history are described. In addition to the friends-of-friends (FOF) catalogues (used exclusively for seeding the Black Hole's in the galaxy formation model), halos and subhaloes are also identified at each redshift by using the publicly available {\sc Ahf}\footnote{{\sc Ahf} is publicly available from  \texttt{http://www.popia.ft.uam.es/AHF}} halo finder \citep{2009ApJS..182..608K}. {\sc Ahf} identifies local over-densities in the adaptively smoothed density field as prospective halo centres. The local potential minimum is computed for each density peak and the particles that are gravitationally bound are identified. The thermal energy of the gas is used in the unbinding procedure by simply adding it to the total specific energy, which is required to be negative. Only peaks with at least 20 bound particles are considered as haloes and retained for further analysis. {\sc Ahf} automatically identifies haloes, subhaloes, sub-subhaloes, etc. 

Magnitudes and colors are computed via a stellar population synthesis model called {\sc Stardust} \citep[see][and references therein for a detailed description]{Devriendt99}. This is used to derive the luminosities from the star particles (i.e. SSP) in each halo. {\sc Stardust} additionally computes the spectral energy distribution from the far-UV to the radio, for an instantaneous starburst of a given mass, age and metallicity. %%%% This is the same as above

Galaxy and halo histories are estimated via merger trees. A tool (aptly named {\sc MergerTree}) is included as part of the {\sc Ahf} package for this purpose. At each redshift accretion events are found by identifying which subhaloes at
a given snapshot are identified as `field' haloes at the previous snapshot. Haloes may also grow via smooth accretion from the environment, namely via gravitationally attracting particles in their vicinity.

Thus the data obtained and retained from the simulation for further analysis consists not just of the raw simulation output, namely the position, velocity, mass and baryonic properties (where applicable) of the particles and cells, but the (sub-)halo finder's {\sc Ahf} catalogue, with bulk baryonic properties computed, as well as merger trees and halo growth data. Throughout this paper only  the {\sc Ahf} post-processing output is used.

\subsection{Identification of Local Groups}
\subsubsection{Low resolution runs}
As explained in Section~\ref{sec:wf} each constrained realisation requires two random seeds. One  source of randomness is associated with the unconstrained regions and the incompleteness of the sampling (including the finite size of the reconstructed box). The second is associated with scales below the reconstruction scale (i.e. non-linear scales). In fact, these two sources of randomness may work together to influence the simulated local volume. To obtain the most accurate ICs suitable for high resolution hydrodynamic simulation,  the problem is therefore approached in a trial and error manner, running many low resolution simulations each with a different combination of large and small scale randomness. These are then searched for the best ICs (according to criteria described below).

In practice close to 1000 DM only, low resolution constrained ICs were generated. These are 147.5~Mpc (100~Mpc$h^{-1}$) periodic boxes filled with 256$^3$ DM particles. A central region of radius $R_{\rm resim} \approx 14.7{\rm~Mpc}~(10{\rm~Mpc}h^{-1})$ is ``zoomed'' in on and populated with the equivalent of $512^3$ effective particles according to the prescription of \cite{1993ApJ...412..455K}. Specifically ICs are generated for the full box with $512^3$ particles. A sphere of radius 14.7 ~Mpc (10 ~Mpc$h^{-1}$) is identified at $z=0$ in the $256^3$ run. This Lagrangian region is tracked back to the ICs. Those high resolution (512$^3$ effective) particles within this volume replace their counterparts in the $256^3$ ICs. Throughout this paper we refer to the resolution level by the effective number of particles in the small high resolution zoom region. The mass resolution achieved is  $m_{\rm DM}=6\times10^{8}{\rm M}_{\odot}$. This is sufficient to search for LGs as the two main haloes will be resolved with a few $\sim10^{3}$ particles. These low-resolution constrained ICs are run to $z=0$, analysed with {\sc AHF} and cosmographically quantified.

The cosmographic quantification attempts to answer the question ``how representative is a given constrained simulation?''. Although there may be differing ideas in the community regarding what constitutes ``representative'', here the focus is on ensuring the simulations are correct (match the data) on large scales and on small scales. On large scales the identification of three robust features of the Local Universe is required: the Virgo cluster, the local void and the local filament. On  small scales, a LG is required. This methodology is similar to that outlined in the ``Local Group Factory'' \citep{2016MNRAS.458..900C}, but slightly more conservative. The algorithm to identify the cosmographic features in the low resolution simulations proceeds with the following criteria:

\begin{figure*}
\includegraphics[width=40pc]{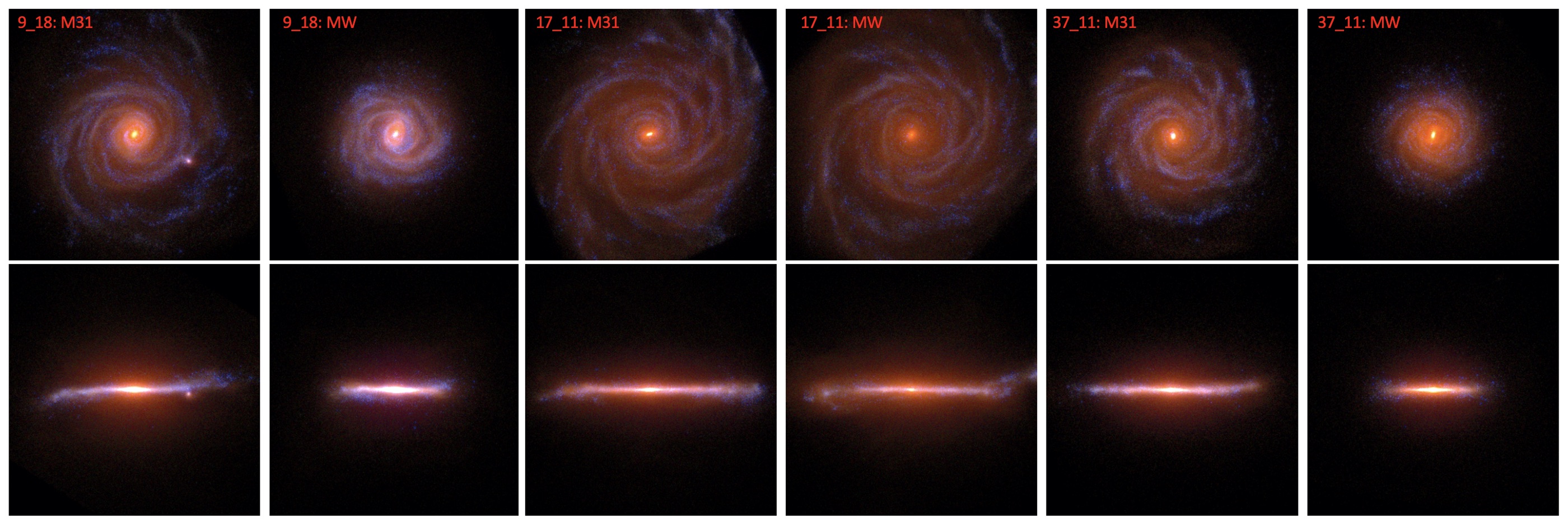}
\caption{Face on (top) and edge on (bottom) mock HST images of the 6 highest resolution {\sc Hestia} LG members. The scale of the box is 30kpc across.}
 \label{fig:hst}
 \end{figure*}

\begin{figure*}
\includegraphics[width=6.5pc]{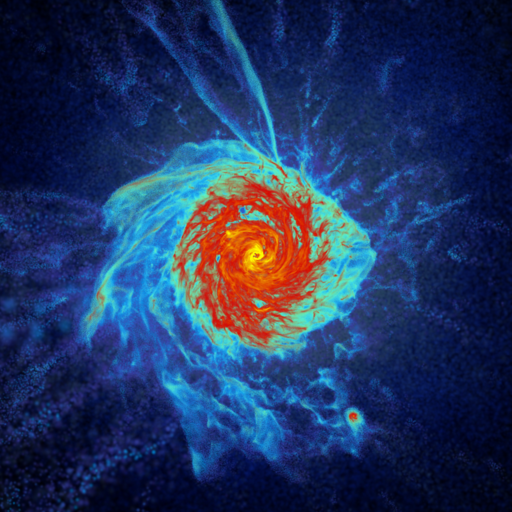}
\includegraphics[width=6.5pc]{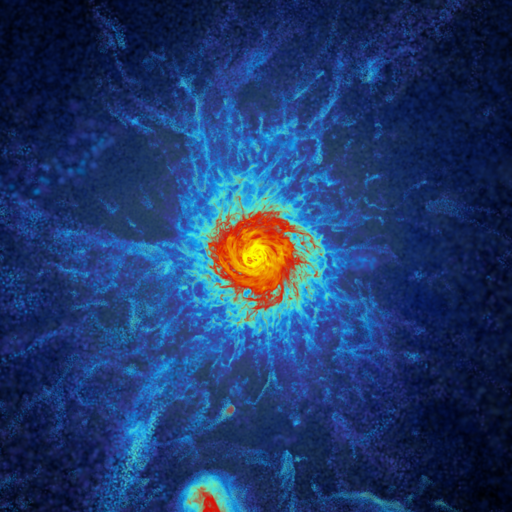}
\reflectbox{\rotatebox[origin=c]{180}{\includegraphics[width=6.5pc]{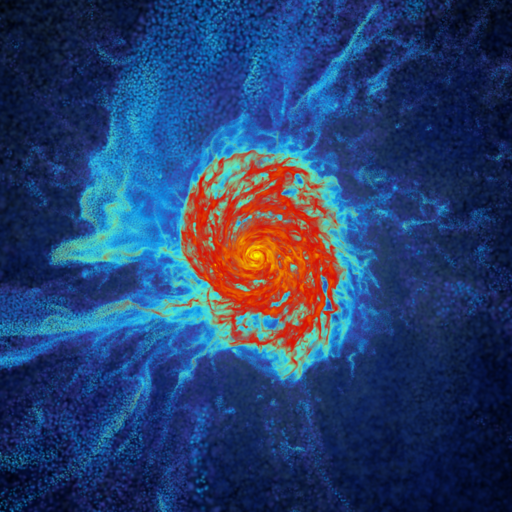}}}
\includegraphics[width=6.5pc]{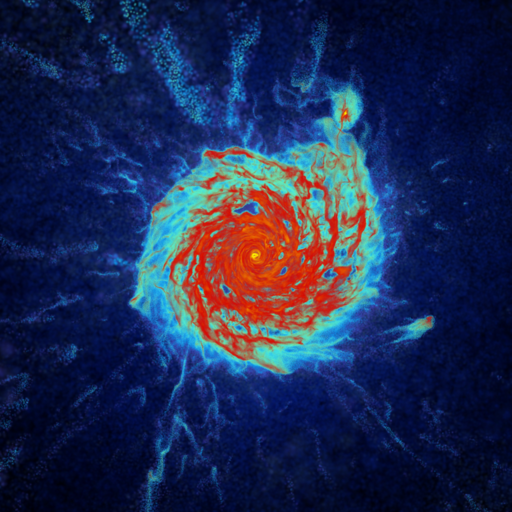}
\includegraphics[width=6.5pc]{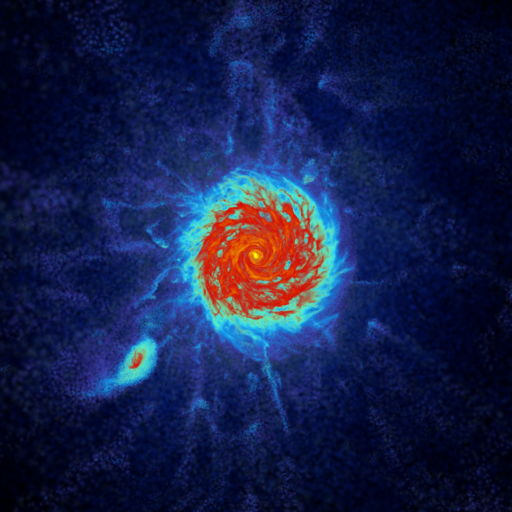}
\includegraphics[width=6.5pc]{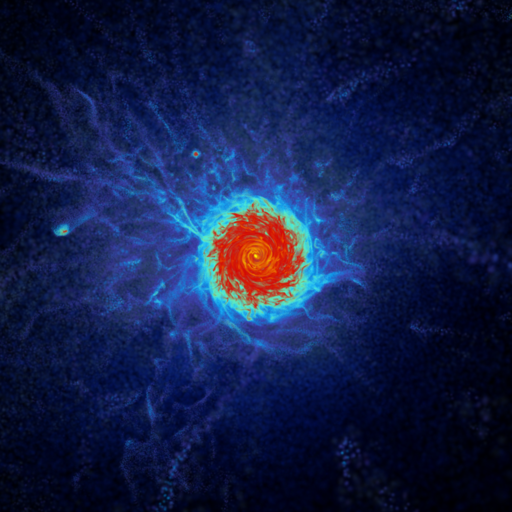}
\includegraphics[width=6.5pc]{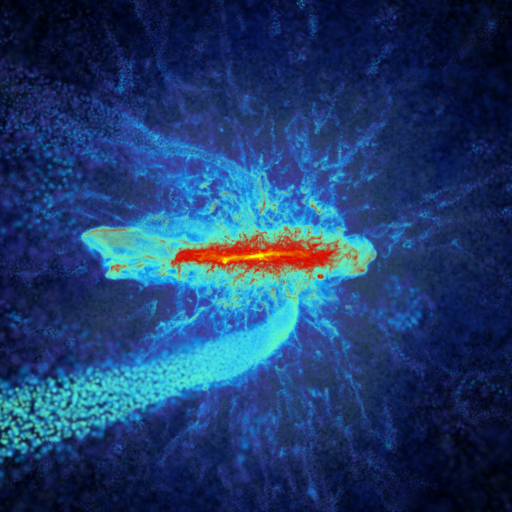}
\includegraphics[width=6.5pc]{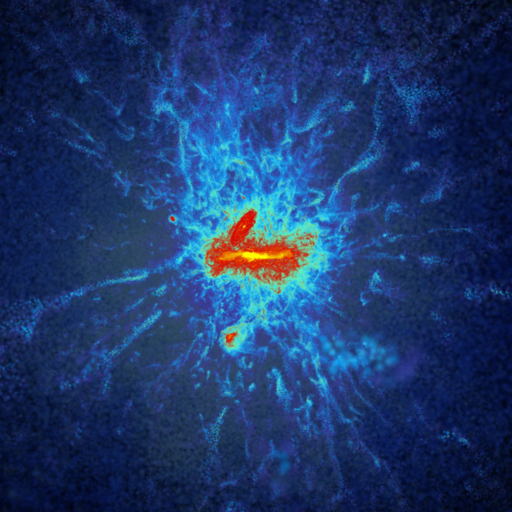}
\includegraphics[width=6.5pc]{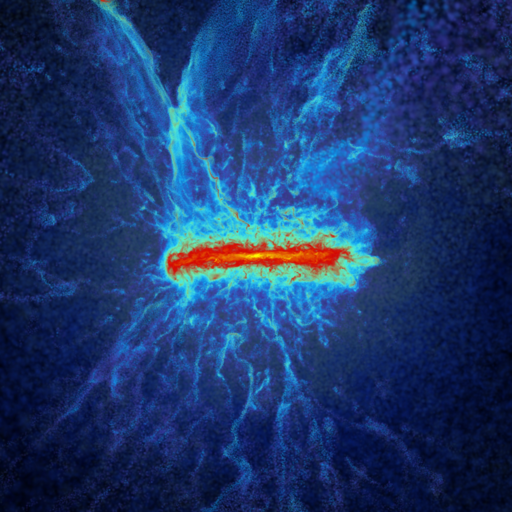}
\includegraphics[width=6.5pc]{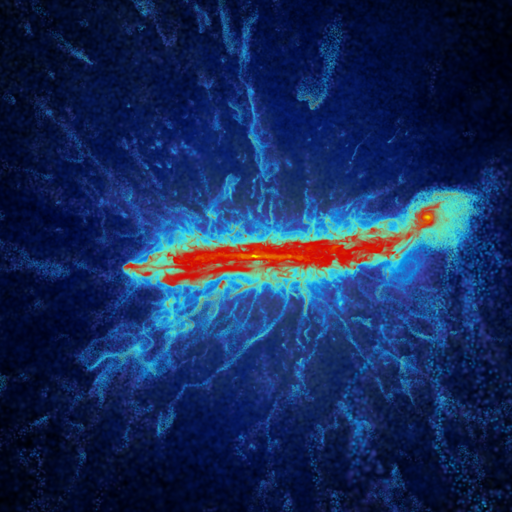}
\includegraphics[width=6.5pc]{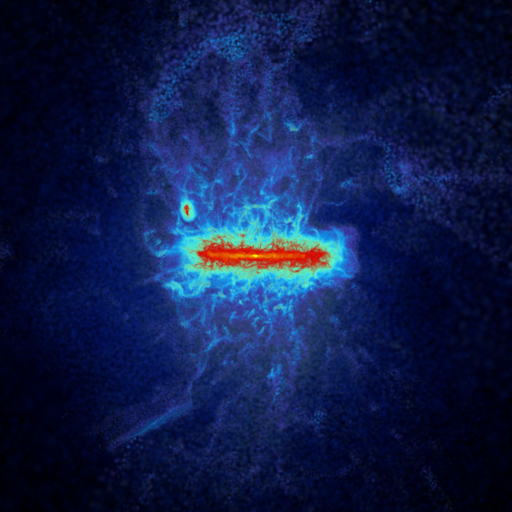}
\includegraphics[width=6.5pc]{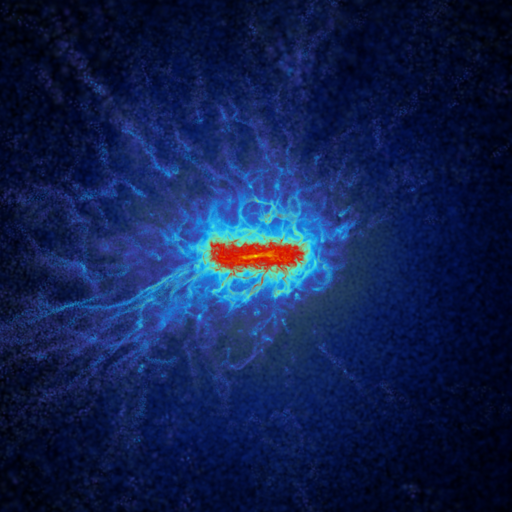}

\caption{Same as Figure~\ref{} but showing the gas density. Red indicates higher gas density, blue-black indicates lower density. The scale of the box is 50 kpc across, highlighting the nearby circum-galactic medium.}
 \label{fig:face-edge-on-gas}
 \end{figure*}

\begin{enumerate}
\item A Virgo cluster mass halo (with mass $> 2\times 10^{14}{\rm M}_{\odot}$) within 7.5~Mpc of where it is expected to form in the simulation, namely at (SGX, SGY, SGZ) = (-3.5,16.0, -0.8) Mpc,  \citep[as per][]{2010MNRAS.405.1075K}. The mass of the observed Virgo cluster is probably a factor of a few greater than the limit of $2\times 10^{14}{\rm M}_{\odot}$ which is the most conservative estimate from \citet{2010MNRAS.405.1075K}. For example the numerical action method of \cite{2017ApJ...850..207S} estimates a mass of 6.3$\pm0.8\times10^{14}{\rm M}_{\odot}$   In practice the Virgo cluster masses adopted here are $3-4\times 10^{14}{\rm M}_{\odot}$ (see Table.~1). The fact that these too are (possibly) smaller than the observed value is due to the small box size of 100~Mpc$h^{-1}$. Constrained simulations performed in larger boxes produce more massive Virgo clusters \citep{2016MNRAS.460.2015S} with little scatter in mass. This is the first criteria - if the simulation lacks a clearly distinguished Virgo cluster with sufficient mass, it is abandoned. This can happen because the reconstruction does not guarantee a virialized halo, but rather an overdensity which may be fragmented in many smaller haloes. 

\item A LG-like pair within 5~Mpc of where it is expected to form in the simulation, namely at the super-galactic origin: (SGX, SGY, SGZ) = (0, 0, 0). These must have:
	\begin{enumerate}
		\item halo mass: $8\times10^{11} < M_{\rm halo}/{\rm M}_{\odot} < 3\times 10^{12}$
		\item separation: $0.5 < d_{\rm sep}/{\rm ~Mpc} < 1.2 $
		\item isolation: no third halo more massive than the smaller one within 2~Mpc of the midpoint
		\item halo mass ratio of smaller to bigger halo $>0.5$
		\item infalling i.e. $v_{\rm rad} < 0$
	\end{enumerate}
\item Since the exact location of a LG is unconstrained, once a LG is identified it is ensured that it is in the correct location with respect to the LSS. Thus the distance from the identified LG to the constrained Virgo cluster is recomputed and required to be within 3.5~Mpc of the true distance.
\item Finally, no other cluster more massive than the simulated Virgo is permitted to exist within a sphere of 20~Mpc centered on the LG, to ensure the primacy of Virgo. With the current observational data and the constraining method there is only a single Virgo-mass cluster which is usually in the correct position. The existence of another cluster in the neighbourhood is extremely rare. Therefore this criterium is essentially obsolete but is kept to be consistent with our previous work. 
\end{enumerate}

A thorough description of the effects of these criteria on the success rate is beyond the scope of this paper. However a few comments are necessary on the restrictiveness of each individual criterium. Although the existence of a large over density at the location of Virgo is a consistent feature of this constrained technique, there is not, necessarily, a large virialized DM halo of mass $>2\times 10^{14}{\rm M}_{\odot}$ there. Namely, there may be many smaller haloes bunched up, or simply a large extended overdensity. This robust feature of the reconstruction and constrained simulation technique consistently produces a local filament, sheet and void that are consistent with observations. A typical constrained simulation is shown in Figure~\ref{fig:CSbox} where the observed location of the Virgo Cluster \citep{2010MNRAS.405.1075K}, the Local Void \citep[from][]{2009AJ....138..323T} and the direction of the Local Filament  \citep[from][]{2015MNRAS.452.1052L} are superimposed on the constrained simulation. Note how well the constrained simulation fits the observational features.

\subsubsection{Intermediate and high resolution hydro runs}
Once the low resolution DM only ICs have been run, their $z=0$ snapshot analysed and the cosmographically successful simulations identified, the ICs are regenerated with a higher resolution of $4096^3$ effective particles  in a 7.4~Mpc (5~Mpc$h^{-1}$) sphere around the center of the LG. Small scale power is added randomly by sampling the $\Lambda$CDM power spectrum at the appropriate scales. These ICs are then run with full MHD and galaxy formation (see Sec.~\ref{sec:hydro}). Gas cells are added by splitting each DM particle into a DM -- gas cell pair with a mass determined by the cosmic baryon fraction. The pair is separated by a distance equal to half the mean inter-particle separation. The centre of mass and its velocity is left unchanged.

The mass and spatial resolution achieved  is $m_{\rm dm}= 1.2\times10^{6}{\rm M}_{\odot}$, $m_{\rm gas}= 1.8\times10^{5}{\rm M}_{\odot}$ and $\epsilon = 340$~pc. At $z=0$ the region uncontaminated by lower-resolution elements is a spherical blob roughly 6~Mpc (4~Mpc$h^{-1}$) in radius that includes the LG and hundreds of field and satellite dwarf galaxies. These simulations represent the intermediate resolution, constrained LG simulations. The properties of the LG members can be found below in Table~\ref{tab:4096runs}.  They constitute the bulk of the simulations analyzed and studied in this paper.

After MHD simulation, the baryonic components of the simulated LGs are examined and subject to additional non-cosmographic criteria, so as to ensure that the simulated galaxies are fair representations of the observed LG. They are visually inspected to ensure they are disky and not bulge dominated.

In order to go to even higher resolution only three ICs are selected (owing to the high computational cost). These are named 09\_18, 17\_11 and 37\_11: the first number refers to the seed for the long waves, while the second refers to the seed for the short waves. Two overlapping 3.7~Mpc (2.5~Mpc$h^{-1}$) spheres are drawn around the two main $z=0$ LG members and are then populated with 8192$^3$ effective particles. The mass and spatial resolution achieved  is $m_{\rm dm}= 1.5\times10^{5}{\rm M}_{\odot}$, $m_{\rm gas}= 2.2\times10^{4}{\rm M}_{\odot}$ and $\epsilon = 220$~pc. Note that there are some slight, expected, more-or-less negligible changes to the LG set up at $z=0$, when run with hydrodynamics at 8192$^3$ resolution compared with 4096$^3$ resolution (see Table 1).

Figure~\ref{fig:hst} and Figure~\ref{fig:face-edge-on-gas} show face and edge on views of the galactic discs produced by these simulations. The orientations are selected according to the eigenvectors of the moment of inertia tensor computed from the stars within 10\% of the virial radius. Figure~\ref{fig:hst}, shows a mock HST image of the stellar component. Such an image is generated by superimposing $U,~V$ and $B$ band luminosities to the Red, Green and Blue colour channels respectively \citep[see][]{2008MNRAS.387...13K}. 
Photometric properties for each star particle can be derived by using stellar population synthesis models to estimate the luminosity in a series of broad-bands. Luminosities for U, V, B, K, g, r, i and z bands from the \cite{2003MNRAS.344.1000B} are tabulated. The effects of dust attenuation are not included\footnote{Note that the method for computing luminosities for the purpose of the mock HST images is not identical to what is used in Figure~\ref{fig:sat_lf}, described in section 3.1. Despite the difference in method, the results are nearly identical. The differences arise from the use of different  pre-existing codes for the purpose of computing luminosities of star particles.}.

\begin{figure*}
 \includegraphics[width=13.333pc]{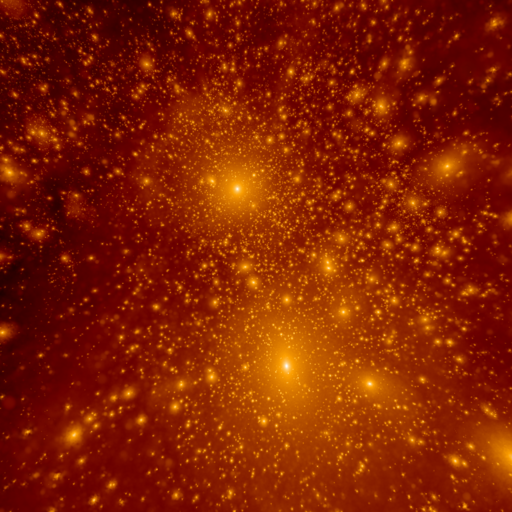}
 \includegraphics[width=13.333pc]{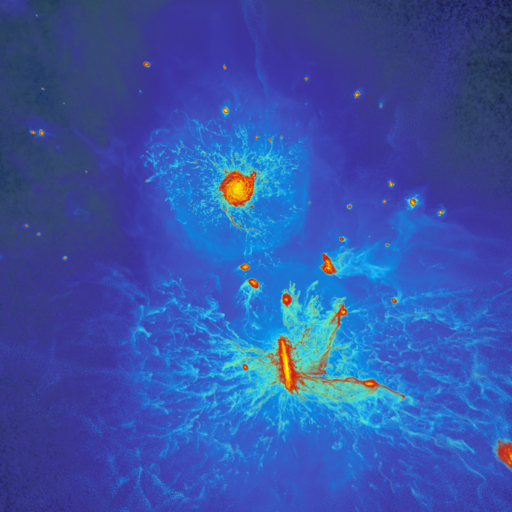}
 \includegraphics[width=13.333pc]{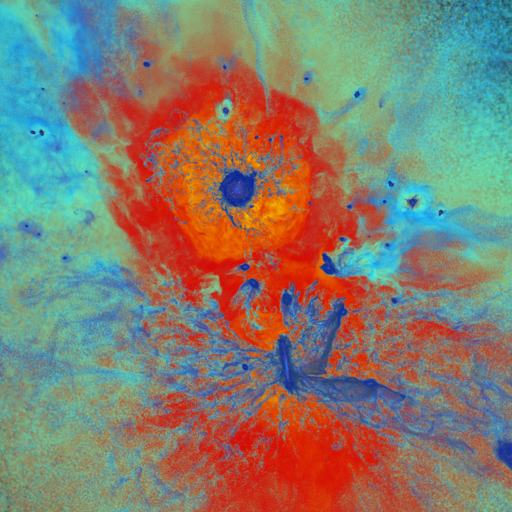}
\caption{One (17\_11) of the high resolution {\sc Hestia} $z=0$ LGs. The panels show from left to right the DM density (white: high density, red-black: low density), gas density (yellow-white: high density, blue-black low density), and density weighted gas temperature (yellow-red: $10^{6}$ K hot gas, dark blue: $10^{4}$ K cold gas). The two main galaxies are clearly seen as is a smattering of accompanying dwarf galaxies and circum-galactic gas. Panels are all centered on the LG and are 3~Mpc across.}
 \label{fig:LG-gas}
 \end{figure*}

Figure~\ref{fig:face-edge-on-gas} shows gas density on a slightly enlarged scale, chosen to highlight circum-galactic gas. Note that, although there are variations in size, mass, bar, etc, most of the {\sc Hestia} LG galaxy members look similar: they are disky spiral galaxies whose halo and stellar mass is within the observed range. See the following sections for a quantification of properties. Figure~\ref{fig:LG-gas} is a zoom out and shows one of the high resolution {\sc Hestia} LGs, including the DM, gas density and density weighted temperature. The thermodynamic interaction between galactic, circum-galactic and inter-galactic LG gas can be seen here.

\begin{table*}
 \centering
 \begin{tabular}{p{0.3cm}p{0.9cm}p{0.9cm}p{0.9cm}lllllllll}
\hline
&&&&&&&&&&&&\\

&ID & $M_{\rm M31}$ & $M_{\rm MW}$ & $M_{\rm MW}/$& $M^{\rm M31}_{\rm star, disc}$&$ M^{\rm MW}_{\rm star, disc} $ & $d_{\rm sep}$ & $v_{\rm rad}$ &$v_{\rm tan}$&$M_{\rm Virgo}$&$d_{\rm Virgo}$&$\Delta_{\rm Virgo}$\\
&&&&$ M_{\rm M31}$&&&&&&&&\\
&&&&&&&&&&&&\\
&&$10^{12}{\rm M}_{\odot}$&$10^{12}{\rm M}_{\odot}$&&$10^{10}{\rm M}_{\odot}$&$10^{10}{\rm M}_{\odot}$&kpc&kms$^{-1}$&kms$^{-1}$ &$10^{14}{\rm M}_{\odot}$&10$^{4}$kpc&Mpc\\
\\\hline
\\
 \parbox[t]{0mm}{\multirow{9}{*}{\rotatebox[origin=l]{90}{intermediate resolution}}} 
 &9\_10   &$  2.06 $&$  1.25 $&$  0.607 $&$  8.13 $&$  4.46 $&$  857 $&$ -61.7 $&$  48.3 $&$  4.28 $&$  1.72 $&$  0.648 $\\
 &9\_16   &$  2.07 $&$  1.25 $&$  0.605 $&$  6.49 $&$  5.87 $&$  752 $&$ -68.8 $&$  59.5 $&$  4.25 $&$  1.73 $&$  0.727 $\\
 &9\_17   &$  2.27 $&$  1.31 $&$  0.577 $&$  7.94 $&$  6.38 $&$  1090 $&$ -10.3 $&$  28.3 $&$  4.30 $&$  1.72 $&$  0.657 $\\
 &9\_18   &$  2.22 $&$  1.92 $&$  0.863 $&$  8.47 $&$  6.41 $&$  856 $&$ -76.1 $&$  55.7 $&$  4.25 $&$  1.71 $&$  0.577 $\\
 &9\_19   &$  2.15 $&$  1.22 $&$  0.566 $&$  7.02 $&$  4.45 $&$  726 $&$ -87.3 $&$  44.8 $&$  4.25 $&$  1.72 $&$  0.639 $\\
&17\_10   &$  2.16 $&$  2.08 $&$  0.961 $&$  7.53 $&$  9.71 $&$  736 $&$ -76.4 $&$  12.1 $&$  3.02 $&$  2.00 $&$  3.42 $\\
&17\_11   &$  2.35 $&$  1.98 $&$  0.845 $&$  11.0 $&$  6.25 $&$  662 $&$ -103.9 $&$  13.9 $&$  3.14 $&$  2.00 $&$  3.43 $\\
&17\_13   &$  2.08 $&$  1.89 $&$  0.910 $&$  6.72 $&$  7.35 $&$  975 $&$ -26.9 $&$  12.6 $&$  3.15 $&$  1.99 $&$  3.29 $\\
&17\_14   &$  2.20 $&$  2.06 $&$  0.938 $&$  11.9 $&$  6.36 $&$  613 $&$ -104.4 $&$  14.1 $&$  3.12 $&$  2.00 $&$  3.43 $\\
&37\_11   &$  1.12 $&$  1.11 $&$  0.993 $&$  4.07 $&$  5.67 $&$  864 $&$ -17.0 $&$  68.8 $&$  3.87 $&$  1.59 $&$  0.680 $\\
&37\_12   &$  1.22 $&$  0.99 $&$  0.812 $&$  5.95 $&$  4.30 $&$  838 $&$ -23.4 $&$  87.5 $&$  3.96 $&$  1.59 $&$  0.646 $\\
&37\_16   &$  1.11 $&$  1.08 $&$  0.973 $&$  4.46 $&$  4.59 $&$  721 $&$ -26.6 $&$  81.4 $&$  3.90 $&$  1.60 $&$  0.593 $\\
& 37\_17   &$    1.23 $&$   0.96 $&$  0.785 $&$    5.17 $&$    4.47 $&$    805 $&$   -41.1 $&$    89.8 $&$    3.96 $&$    1.59 $&$    0.693 $\\
&&&&&&&&&\\
\hline 
 \parbox[t]{8mm}{\multirow{7}{*}{\rotatebox[origin=c]{90}{~~~~~~~~~~high res}}} \\
& 9\_18    &$  2.13 $&$  1.94 $&$  0.913 $&$  11.4 $&$  7.70 $&$  866 $&$ -74.0 $&$  54.0 $&$  4.26 $&$  1.71 $&$  0.565 $\\
&17\_11   &$  2.30 $&$  1.96 $&$  0.852 $&$  9.90 $&$  8.38 $&$  675 $&$ -102.2 $&$  137 $ &3.11&2.00&3.44\\
&37\_11   &$  1.09 $&$  1.04 $&$  0.980 $&$  4.12 $&$  5.70 $&$  850 $&$ -8.86 $&$  71.1 $& 3.82&1.57&0.867\\
&&&&&&&&&\\
\hline 
\\
Observation&&0.6-2.0 &1.0-2.1&  0.5 -- 1 & 8 -- 15 & 5 -- 10 & 785$\pm 25$ &  -110 &0 -- 60&1-8&1.7\\
References&&[1,2,3,4]& [5,6,7,8,9]&&[2,10,11]&[11,12,13]&[14]&[15]&[16,17,18]&[19,20,21]&[20]\\
&&& &&&&&&&[22,23]&\\
&&&&&&&&&\\
\hline 
  \hline
 \end{tabular}
 \caption{ The $z=0$ properties of the intermediate resolution and high resolution runs. We adopt the convention that the more massive halo is termed M31 while the less massive one is termed the MW. $M_{\rm halo}$ is defined as the mass within a radius that encloses a mean density equal to 200$\rho_{\rm crit}$. From left column to right: the simulation ID; $M_{\rm M31}$: the virial mass of the simulated M31 halo $[10^{12}{\rm M}_{\odot}]$;  $M_{\rm MW}$: the virial mass of the simulated MW halo $[10^{12}{\rm M}_{\odot}]$; $M_{\rm MW}$/$M_{\rm M31}$: the mass ratio of the two; $M^{\rm M31}_{\rm star, disc}$: the mass of the stellar disc of M31 (determined as the stars within 0.15$r_{\rm vir}$) $[10^{10}{\rm M}_{\odot}]$; $M^{\rm MW}_{\rm star, disc}$: the mass of the stellar disc of the MW $[10^{10}{\rm M}_{\odot}]$; $d_{\rm sep}$: the separation of the two halos [kpc]; $v_{\rm rad}$: the radial velocity (negative $v_{\rm rad}$ indicates in-falling motion) [km s$^{-1}$];  $v_{\rm tan}$: the tangential velocity [km~s$^{-1}$]; $M_{\rm Virgo}$: the mass of the Virgo Cluster $[10^{14}{\rm M}_{\odot}]$; $d_{\rm Virgo}$: the distance to the Virgo cluster from the LG [$10^{4}$~kpc]; $\Delta_{\rm Virgo}$: the distance the simulated Virgo cluster is from the observed one [Mpc].  References [1]: \citet{2018MNRAS.475.4043K}; [2] \citet{2012A&A...546A...4T}; [3] \citet{2014MNRAS.443.1688D}; [4] \citet{2010A&A...511A..89C}; [5] \citet{2019A&A...621A..56P}; [6] \citet{2018ApJ...866..121H}; [7] \citet{2018A&A...616L...9M}; [8] \citet{2019ApJ...873..118W}; [9] \citet{2017MNRAS.465.3724Z}; [10] \citet{2014AJ....147..109S}; [11] \citet{2017MNRAS.465...76M}; [12]  \citet{2015ApJ...806...96L}; [13] \citet{2014ApJ...794...59K}; [14]\citet{2012AJ....144....4M}; [15]\citet{2006Ap.....49....3K}; [16] \citet{2008ApJ...678..187V}; [17] \citet{2012ApJ...753....8V}; [18] \citet{2019ApJ...872...24V}, [19] \citet{2011A&A...532A.104N}, [20] \citet{2010MNRAS.405.1075K}, [21] \citet{2014ApJ...782....4K}, [22] \citet{2017ApJ...850..207S}, [23] \citet{2017MNRAS.469.1476S}.}
\label{tab:4096runs}
\end{table*}

\section{Results}
The presentation of the results is split into two parts: first a description of the properties of the two main simulated LGs members in comparison to the observed values is presented. The subsequent section focuses on a description of the satellite dwarf galaxies in the simulated LG. This pedagogy is motivated by a qualitative and quantitative comparison with the observed LG. Namely our comparison is limited to properties of the LG where there is good reliable observational data. Characteristics that cannot be directly compared with observations are omitted.

\subsection{A description of the main {\sc Hestia} LG members}
\subsubsection{Local Group dynamics and mass accretion history}

Before a description of how the simulated LGs compare to the observed LG in terms of their internal (namely baryonic) properties is shown, in the following section their bulk features - dynamics and mass - are presented together with those of the observed LG. It must be emphasised that the LGs in this work have been ``bred'' to have certain properties consistent with observations - they have been constructed and selected so as to be accurate representations of the observed LG.

\begin{figure}
 \includegraphics[width=20pc]{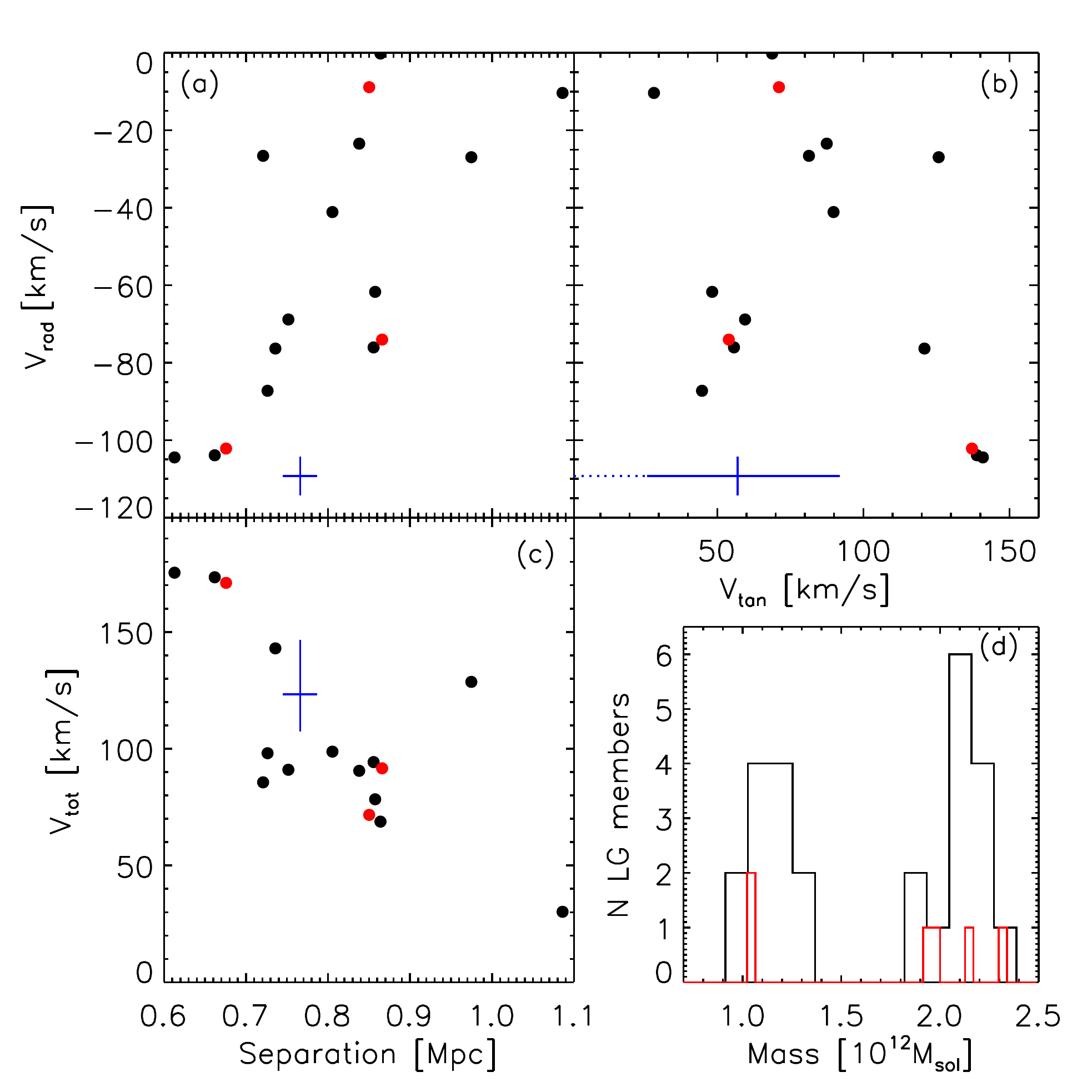}
\caption{The dynamics of simulated LGs.  (a): The relative radial velocity versus pair separation.  (b): Radial versus tangential velocity.  (c): The total relative velocity versus pair separation. (d): The mass function for all LG members. The black dots/line refer to LGs simulated at the intermediate resolution while the red dots/line are for the high  resolution. In blue we show the observational values with error bars. Solid blue line is $v_{\rm tan}=57\pm30$ kms$^{-1}$ from Gaia proper motions (\citealt{2019ApJ...872...24V}), while dotted line is $v_{\rm tan} <34$kms$^{-1}$ from HST proper motions (\citealt{2012ApJ...753....8V}).}
 \label{fig:mpairs}
 \end{figure}

 Figure~\ref{fig:mpairs} focuses on the dynamics of the pair, showing how the relative radial and tangential velocities and pair separation are related and compared with the observed values. Note that although {\sc Hestia} LGs are not able to perfectly (within the errors) match simultaneously the radial velocity, the tangential velocity {\it and} the separation as observed in the LG, they are fairly close representations. Recall that by construction they must have separations of 500-1,200 kpc and negative radial velocities (no requirement is made of the tangential velocity). For example, the systems shown here, have relative radial velocities of $< -100$~kms$^{-1}$ and have tangential velocities of $\sim 150$~kms$^{-1}$ which is greater than the observed value of $\sim 20-70$~kms$^{-1}$ from proper motions \citep{2008ApJ...678..187V,2012ApJ...753....8V,2018A&A...619A.103F}, although in agreement with the more controversial value of $\sim 127$~kms$^{-1}$ from satellite orbital modelling \citep{2016MNRAS.456.4432S}. Similarly, most of the systems that have radial velocities consistent with the data ($\sim - 110$~kms$^{-1}$), have separations of $<700$~kpc, slightly smaller than the observed value of 785~kpc. The total relative velocity (defined as the square root of the sum of the squares of the tangential and radial velocities) and separation is, however, consistent with that of the LGs. 

\begin{figure}
 \includegraphics[width=20pc]{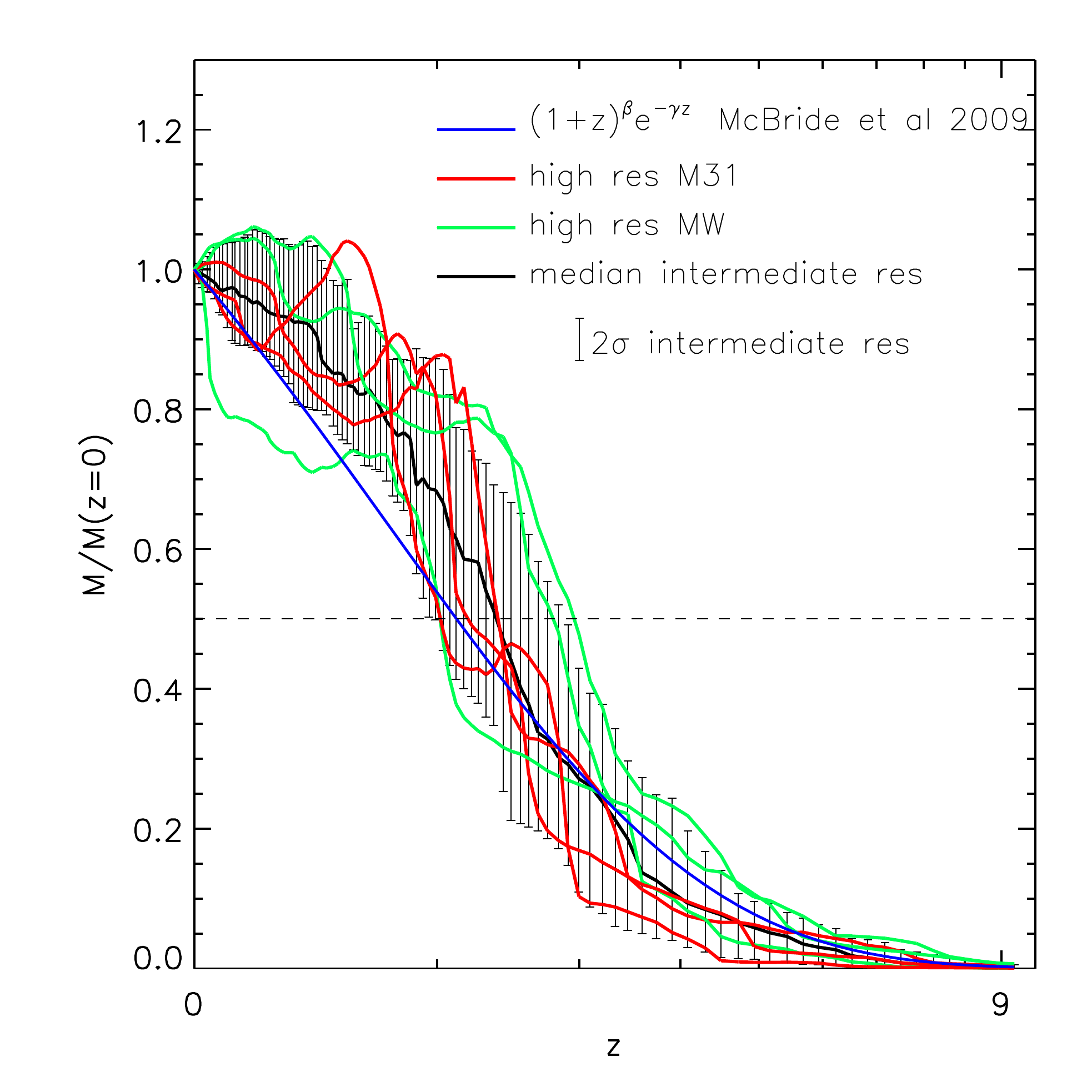}
\caption{The mass accretion history of the largest progenitor halo of the two main LG haloes. The growth history of each $z=0$ LG halo is computed via a merger tree described in Sec.~\ref{sec:pp}. The intermediate and high resolution runs are indicated by the black and red lines, respectively. In blue an empirical relation obtained by \citealt{2009MNRAS.398.1858M} by averaging over many MW/M31 mass haloes drawn from a large cosmological simulation is shown. For the most part, {\sc Hestia} LGs grow slower or at the same rate as their unconstrained counterparts at early times (say at $z>2$) but then have a growth spurt (at around $1<z<2$) that results in them having already assembled the bulk of their mass before their unconstrained analogs.}
 \label{fig:mah}
 \end{figure}

  The mass accretion history of the {\sc Hestia} LGs is shown in Figure~\ref{fig:mah} and computed by linking the most massive progenitors between snapshots (see Section~\ref{sec:pp}). The mass accretion histories of the 6 high-resolution LG members are shown as individual red and green curves (for M31 and MW respectively), while the mass accretion histories of the 26 intermediate resolution LG galaxies are quantified in terms of their median and 2$\sigma$ spread. The mass accretion  histories are compared with an empirical curve suggested by \cite{2009MNRAS.398.1858M} obtained by averaging over the accretion history of many dark matter haloes with masses similar to the MW drawn from a large cosmological simulation without environment consideration. Although this is not an observed quantity, it is included here to highlight a key difference between the haloes in our (constrained) simulations and those found in unconstrained simulations of the same mass. The empirical \cite{2009MNRAS.398.1858M} fit allows us to compare if there are any systematic differences introduced by the combination of constrained ICs and selection criteria. Indeed this is the case:  {\sc Hestia} LGs grow slower or at the same rate as their unconstrained counterparts at early times (at $z\lsim2$). {\sc Hestia} LGs appear to have a ``growth spurt'' (between $1\lsim z\lsim2$) where they increase their assembled mass from $\sim 10-40\%$ such that  $\sim 70-90\%$ of their mass is in place by $z\approx 1$. This can also be seen by the redshift at which the haloes have acquired half their $z=0$ mass: for unconstrained similarly massed MW/M31 analogs this occurs, on average, at $z = 1.1$. While for {\sc Hestia} LGs this occurs, on average, at $z=1.4$ roughly one billion years earlier. Specifically, only a fifth of {\sc Hestia} MW/M31 members assemble half their mass later than average unconstrained similarly massed halo. Note that a few of the Auriga haloes show similar growth spurts \citep{2019MNRAS.484.4471F}, although these are not controlled for cosmographic environment. Also, although not shown here, the median mass accretion history of the Auriga suite is coincident with the \cite{2009MNRAS.398.1858M} curve for $z<1$ and roughly follows it for $z>2$, suggesting that the mass accretion history of the {\sc Hestia} simulations is directly influenced by the constrained nature of the environment.

An inspection of the mass accretion history of Figure~\ref{fig:mah} highlights the salient point that to have morphologically similar galaxies to the observed LG, the mass accretion history must be systematically different from equally massed, but environmentally unconstrained, analogs \citep[e.g. see][]{2012MNRAS.423.1544S,2006MNRAS.366..899N}. In fact, a systematic deviation of around 1$\sigma$ between the mean mass accretion history of constrained and unconstrained LGs, was found by \cite{2019MNRAS.tmp.2688C}. Note that most simulations of ``MW type galaxies'' are not random but require either quiet merger histories or isolation. What is perhaps surprising is that constrained simulations based on measurement of the local velocity field  has already been shown to produce significantly and systemically different mass accretion histories for the simulated Virgo cluster \citep{2016MNRAS.460.2015S,2019MNRAS.486.3951S}. Here a similar behaviour is seen on mass scales that are 2 orders of magnitude smaller.

\subsubsection{Baryonic components of individual LG galaxies}
\begin{figure}
 \includegraphics[width=20pc]{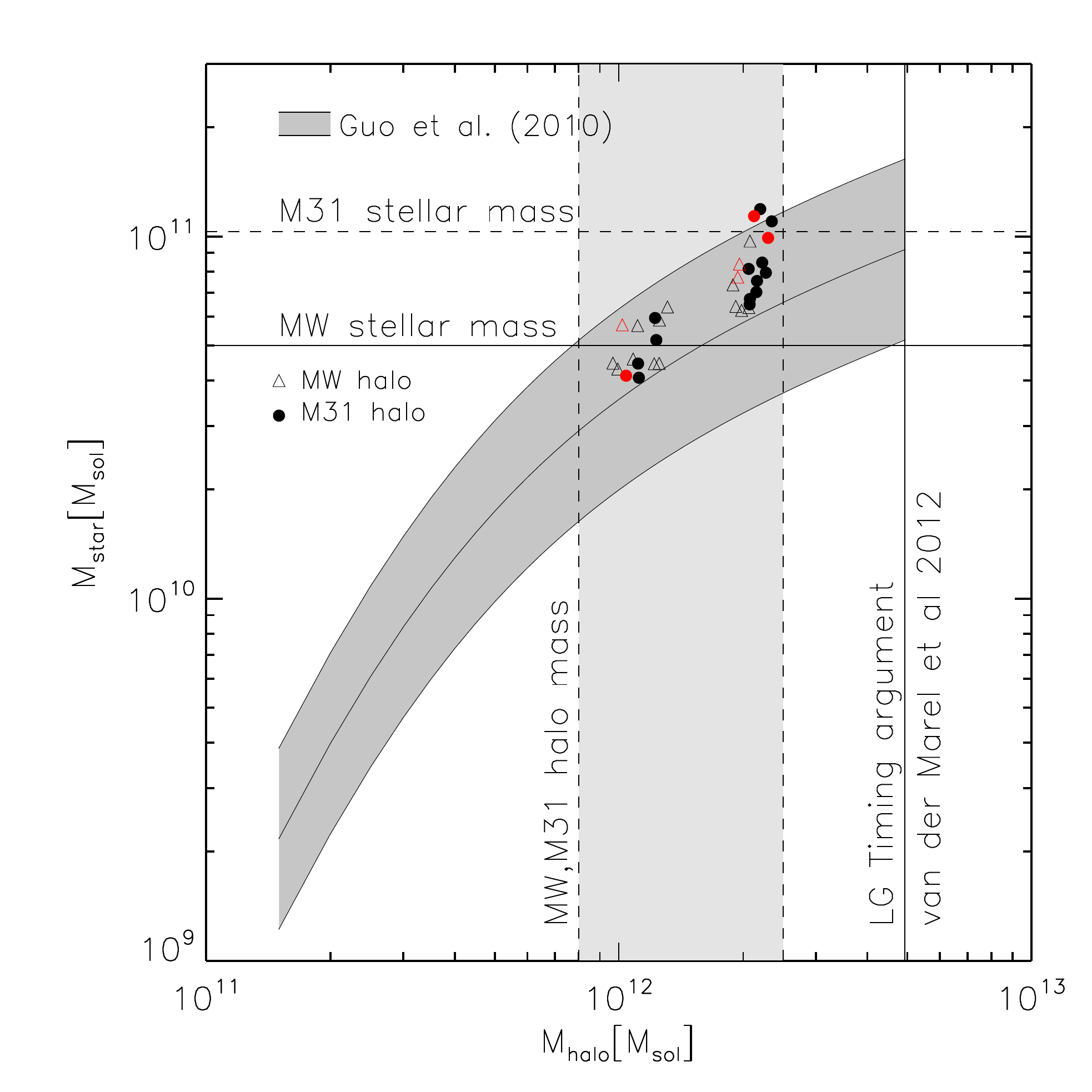}
\caption{The $M_{\rm star}$ vs $M_{\rm halo}$ relation for the {\sc Hestia} LGs. Plotted in black is each galaxy at the intermediate ($4096^3$) resolution while in red is the high ($8192^3$) resolution. In dark grey we plot the $M_{\rm star} - M_{\rm halo}$ relation suggested by \citet{2010MNRAS.404.1111G}. Also included are constraints on the MW and M31 dark halo and stellar mass  (see Table.~\ref{tab:4096runs}), as well as the total LG mass from the timing argument, based on measurements of M31's proper motion \citep{2012ApJ...753....8V}}
 \label{fig:MstarMhalo}
 \end{figure}

In this section the baryonic component of the simulated LG members is examined. Figure~\ref{fig:MstarMhalo}  presents the $M_{\rm star}$ vs. $M_{\rm halo}$ relation for the simulated LGs, along with the empirical $\Lambda$CDM relationship suggested by \cite{2010MNRAS.404.1111G}. Note that the sample sits well within both the permitted range suggested by \cite{2010MNRAS.404.1111G}, as well as the range consistent with observations. It is thus noted that the location of a galaxy-halo pair on such a plot is influenced principally by the Auriga galaxy formation model rather than the constraining nature of the initial conditions. These simulated LGs are also fully consistent with the upper most value for the mass of the entire LG inferred from the timing argument \citep[from proper motion measurements of M31 by][see also \citealt{2017JCAP...12..034M}]{2012ApJ...753....8V}. Note however, that due to the still small number of realizations a continuous distribution of halo masses is not produced, specifically lacking haloes with $M_{\rm halo}\approx 1.25-1.9 \times 10^{12}M_{\odot}$, which would be the median of the observationally permitted limits. This shortcoming is entirely due to the (relatively) low number of LGs produced and examined here, and not to a physically meaningful reason. (This gap is also visible in Figure~\ref{fig:mpairs}d). 

\begin{figure*}
 \includegraphics[width=42pc]{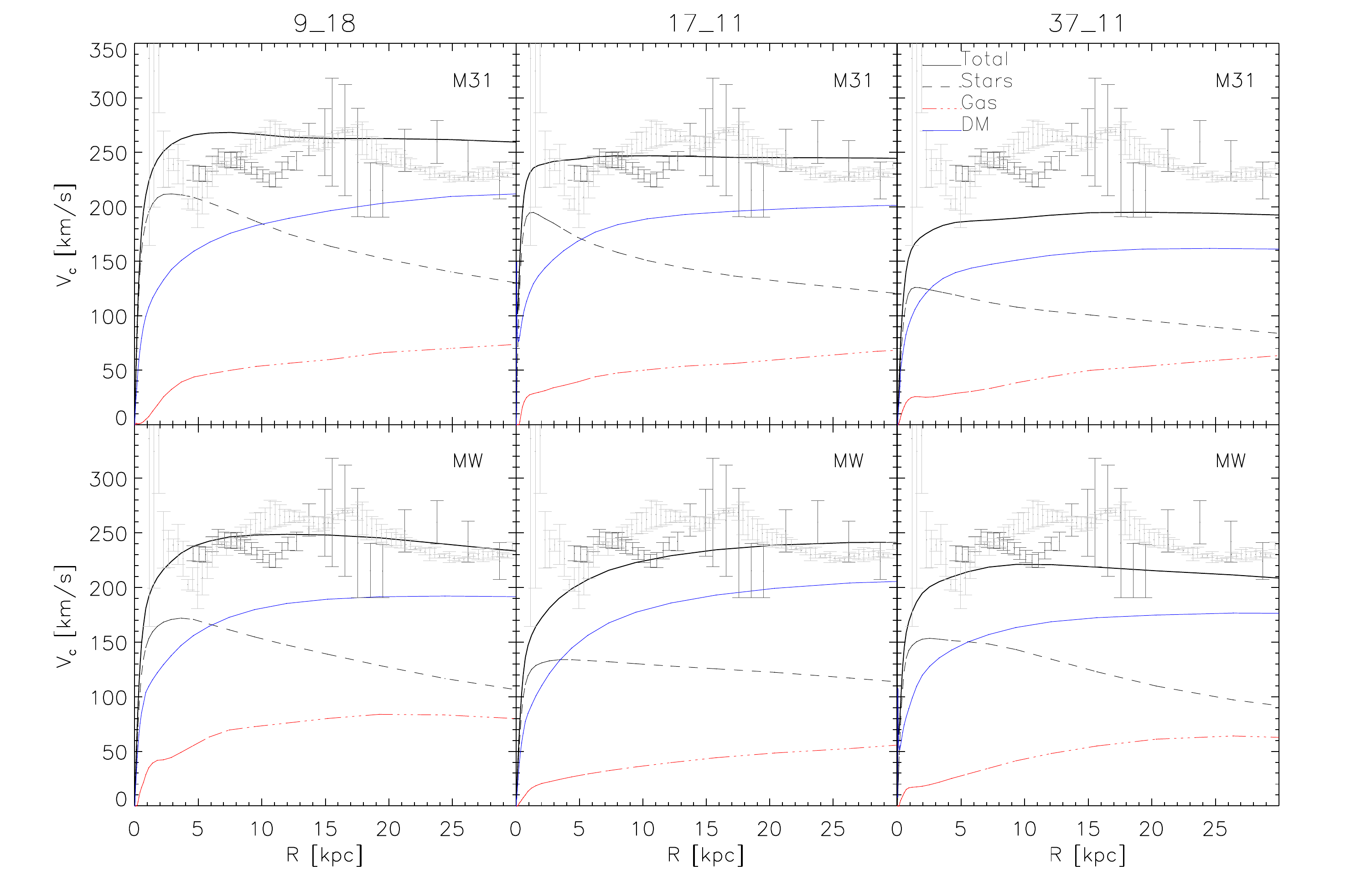}
\caption{Circular Velocity of the high resolution simulated LGs considered here. The more massive halo (M31 analog) is plotted in the upper row, while the less massive one (MW analog) is shown in the bottom row. The total circular velocity is drawn as the black line while the circular velocity for the stellar, gaseous, and dark matter  components is shown as dashed black, red, and blue, respectively. Observational values for the MW from \citealt{2016MNRAS.463.2623H} and for M31 from \citealt{2009ApJ...705.1395C} are shown as the black and grey error bars, respectively. Note that within 30kpc, $V_{\rm c}^{\rm MW}<V_{\rm c}^{\rm M31}$, consistent with the higher stellar mass shown in Table~1.}
 \label{fig:vcirc}
 \end{figure*}

The individual LG galaxies simulated here have circular velocity profiles that match well the observational data. Specific features of a galaxy's circular velocity profile - a steep rise in the inner parts (within $\sim$ kpc) followed by a flattening in the outer parts, are generic features of how disc galaxies form in this model of the $\Lambda$CDM paradigm. Figure~\ref{fig:vcirc} shows the circular velocity profiles computed from the mass distribution namely $v_{\rm circ}(r)=\sqrt{2GM(<r)/r}$, where $M$ is stellar mass, gas mass, dark matter mass or total mass. The observations of MW's and M31's circular velocity  published by \cite{2016MNRAS.463.2623H} and \cite{2009ApJ...705.1395C} respectively, are shown as well. Indeed all circular velocity profiles show this observed behavior, and are included as a dynamical sanity check of the kinematics of these galaxies. 

\subsubsection{Morphology of LG galaxies}
The stellar component of the simulated LG can be decomposed into a bulge and disk component by fitting an exponential disk and bulge to the estimated surface brightness profile. Accordingly, an effective bulge radius ($R_{\rm eff}$) and disc scale length ($R_{\rm disc}$) can be inferred. Note that this is different from a dynamical bulge-disc decomposition often employed in numerical simulations. The reason this dynamical measure is avoided here is because observationally such decompositions are unobtainable, while surface brightness modelling can be performed on both the MW \citep[e.g. see][for a review]{2016ARA&A..54..529B} and on M31 \citep{2011ApJ...739...20C} amongst other galaxies.

\begin{figure*}

 \includegraphics[width=42pc]{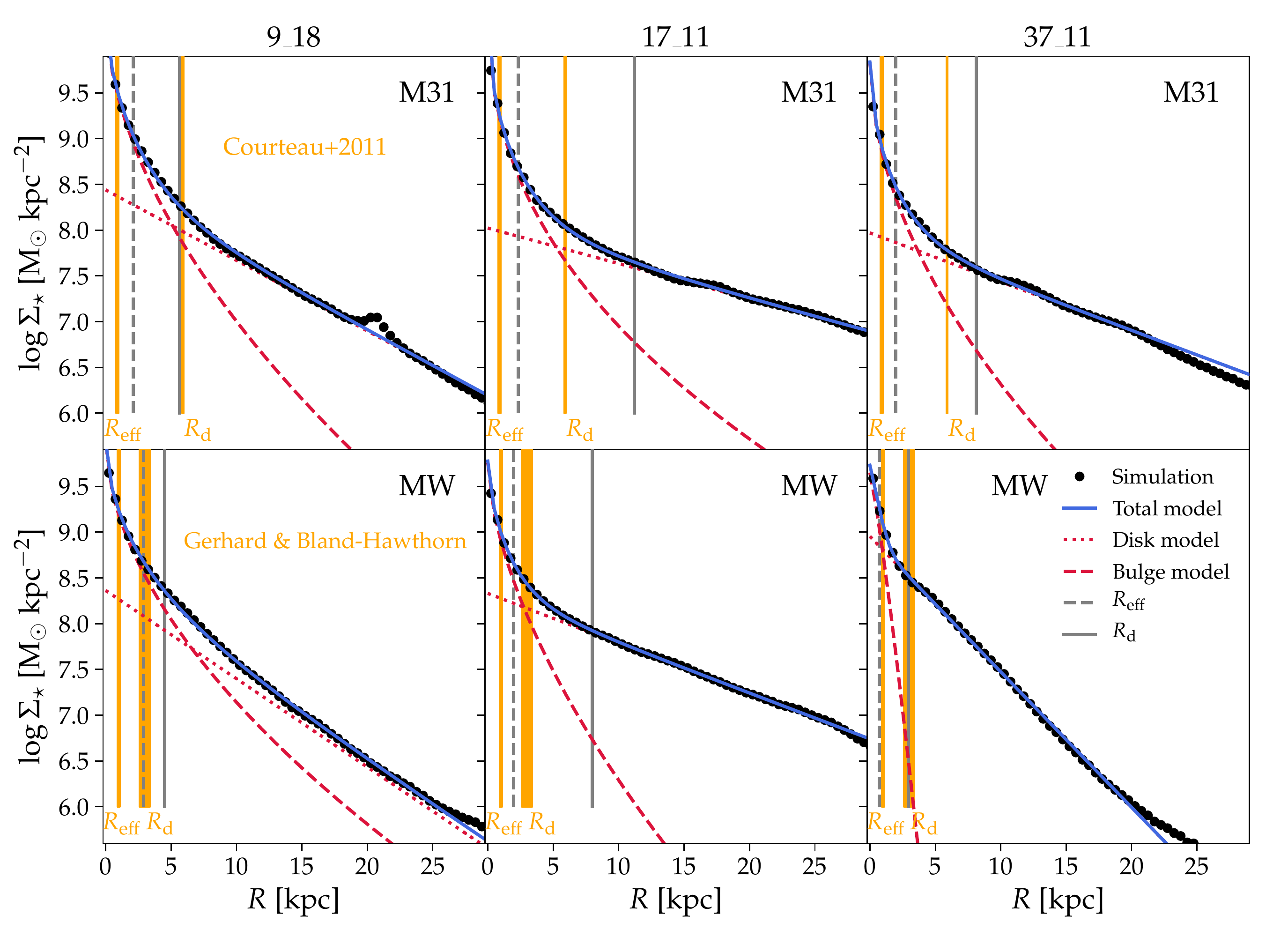}
\caption{The surface brightness profile of the simulated LG galaxies (M31 top row, MW bottom row) is modelled by the sum of a S\'ersic profile describing the bulge (dashed red line) and an exponentially decaying disk (dotted red line). The total model is shown by the blue curve, while the simulation's values are black dots. The bulge effective radius ($R_{\rm eff}$) and the disk scale length ($R_{\rm d}$) for the simulation are shown with dashed and solid grey vertical lines, while the observed values for M31 and the MW are shown in orange (with the width of the bands indicating the observational error). See Table~\ref{Table:StructureFitting}.}
 \label{fig:surf}
 \end{figure*}

Figure ~\ref{fig:surf}  shows the stellar surface density profile of the simulated LG galaxies, as seen from a face-on projection. The method presented in \cite{2017MNRAS.470.3946S} is employed wherein the contribution of a bulge and disk component is modelled with a S\'ersic and an exponential profile, respectively. The best-fitting parameters have been obtained in a simultaneous fit of the disk and bulge. Only  stars within $0.15R_{200}$ are included in the fit to avoid a contaminating contribution from halo stars. The fitting parameters are presented in Table~\ref{Table:StructureFitting} along with the estimated mass belonging to the  bulge and disk ($D/T$). 

Although a robust disc-bulge decomposition of the Galaxy is complicated by our location within it as well as the overlap of these two components, estimates can be made. The situation for M31 is easier, given our observational vantage point. Two salient points may be inferred from Figure~\ref{fig:surf}: the LG members in these constrained simulation have fairly well-defined bulge and disc components and these are not inconsistent with the published observational values. In some cases, for example the simulated M31 in simulation 09\_18, the values are very close to the observations. Similarly the values for the MW in simulation 37\_11 are in good agreement with the observations. The conclusions one may draw from such a finding is that even by this specific, highly non-linear metric, the simulations presented here resemble the true LG morphologically.

\begin{table*}
\centering
%\begin{minipage}{.53\textwidth}
\begin{tabular}{ccccccccc}
\hline\hline
\\
ID & $R^{\rm M31}_\text{eff}$ & $R^{\rm MW}_\text{eff}$ & $R^{\rm M31}_\text{d}$ & $R^{\rm MW}_\text{d}$& $n_{\rm M31}$  & $n_{\rm MW}$ & $D/T_{\rm M31}$ & $D/T_{\rm MW}$\\
\\
& kpc & kpc&kpc &kpc&&&&\\
\\
\hline
\\
9\_18 & 2.11 &2.88&  5.64 &4.50& 1.81&1.80 &0.46 & 0.38\\
17\_11&2.32& 1.95 &11.21 &7.97&2.28&1.64&0.64&0.82\\
37\_11 &1.99&0.71&8.13&2.93&1.87&1.81&0.72&0.88\\
\\
\hline
\\
Observations&1.0$\pm$0.2&$\sim1$&5.3$\pm$0.5&2.5$\pm$0.5&2.2$\pm0.3$&--&--&--\\
References&[1]&[2]&[1]&[3]&[1]\\
\\

\hline\hline
\end{tabular}
\caption{Structural parameters for the high resolution M31 and MW. The columns show galaxy  simulation  ID,  effective radius of the S\'ersic profile describing the bulge (in kpc) for M31 and MW, the disc scale length (in kpc) for M31 and the MW, the S\'ersic index for M31 and MW,  and the disc-to-total mass fraction for M31 and MW. References: [1] \citealt{2011ApJ...739...20C}; [2] Gerhard \& Bland-Hawthorn (private communication); [3] \citealt{2016ARA&A..54..529B}}
\label{Table:StructureFitting}
\end{table*}

\subsection{A description of the simulated LG dwarfs}
The simulations presented here allow for the study of both LG satellites galaxies, as well as the ``field dwarfs'' in the LG. Only the former are shown here, leaving a description of the field dwarfs to further study. This is because satellite dwarfs can be easily defined (namely substructures within a virial radius) while a complete sample of field dwarfs of the LG is not as straightforward. This is further complicated by the complex (and individual) topology of the resimulated high resolution region.

\subsubsection{Mass and Luminosity function of satellite dwarfs}
\begin{figure}
 \includegraphics[width=20pc]{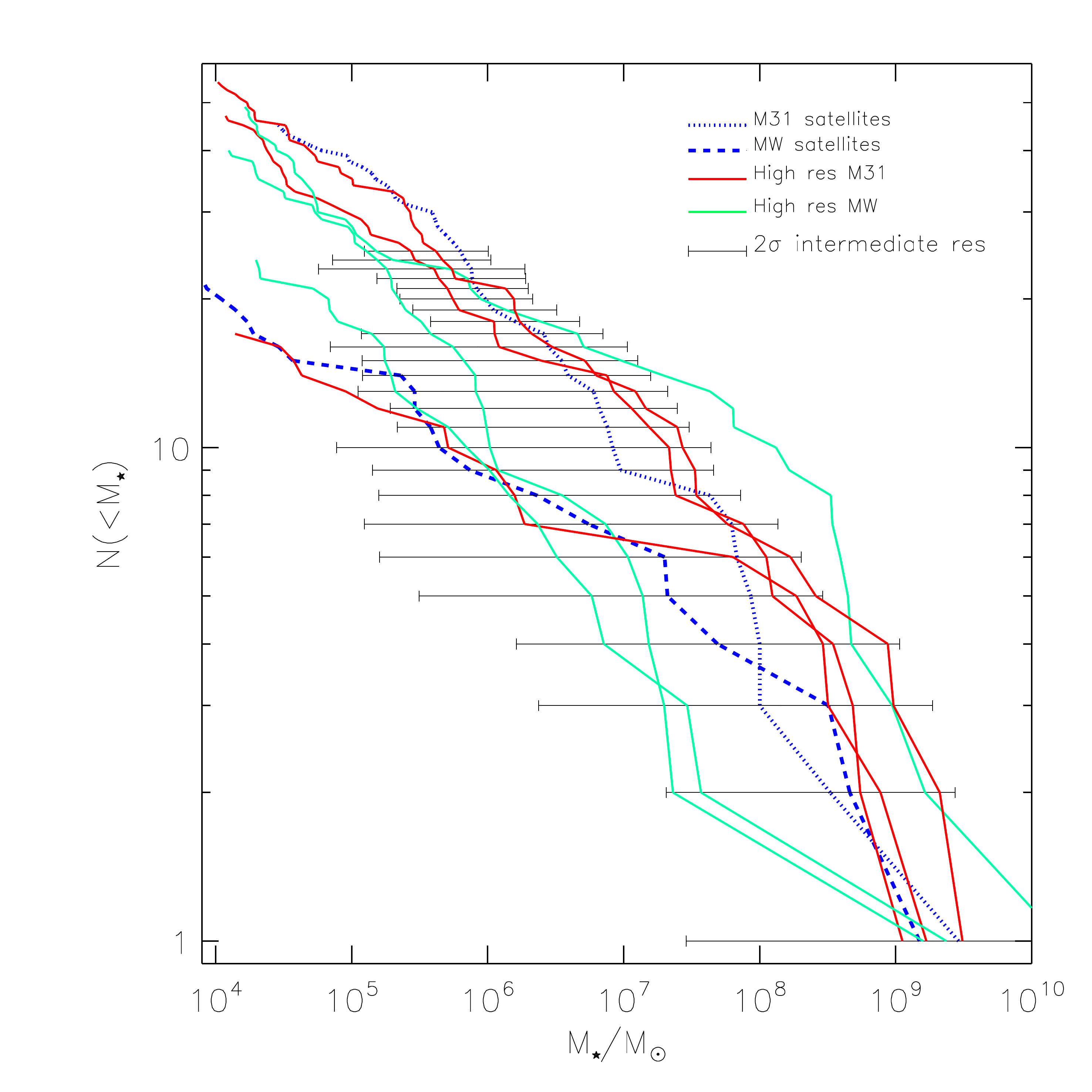}
\caption{The stellar mass function of satellite galaxies in the {\sc Hestia} runs. The intermediate resolution simulations are shown as 2$\sigma$ spreads in black, while the high resolution simulations are shown individually in red and green. In blue dotted and dashed lines we show the observed stellar mass function for the satellites of M31 and the MW, respectively. For the observed data, stellar masses have been inferred by applying  \citealt{2008MNRAS.390.1453W} mass-to-light ratios to the luminosities published in \citealt{2012AJ....144....4M}. Note that the observations are likely to be incomplete below $M_{\star} < 10^{5} M_{\odot}$.}
 \label{fig:sat_mf}
 \end{figure}

One of the most basic tests of any theoretical model which attempts to reproduce the LG, is its prediction of the distribution of satellite galaxy mass.   Only substructures that are well resolved with more than 100 star particles and within 250kpc are shown here. Figure \ref{fig:sat_mf} shows such simulated stellar mass functions. Each of the 6 high resolution LG members are shown individually in red and green (for M31 and MW respectively), while stellar mass functions for the intermediate resolution are grouped together and their $2\sigma$ spread is shown. The width of this spread indicates that there is considerable variance in the satellite mass functions: the most massive satellite in one LG has a mass of a mere few $10^{7}M_{\odot}$, while in others the most massive satellite can be up to three orders of magnitude larger. In the highest resolution runs a difference in the number of satellites (most likely owing to the factor of $\sim$2 differences in LG  masses) is seen: in one case there are more than 50 satellites (more massive than $M_{\star}>10^{4}M_{\odot}$) in another just 15.

\begin{figure}
 \includegraphics[width=20pc]{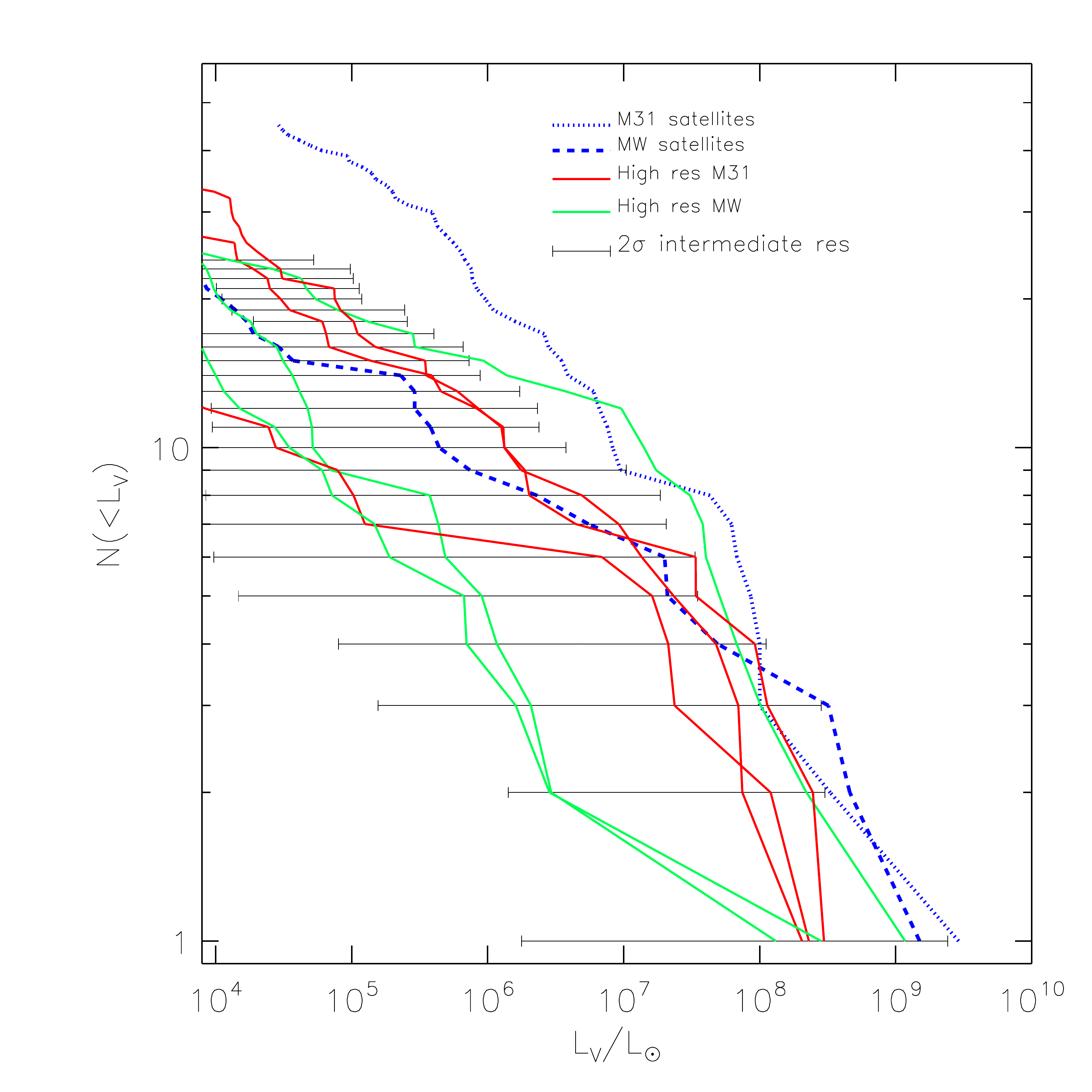}
\caption{The $V$-band luminosity function of satellite galaxies in the {\sc Hestia} runs. The intermediate resolution simulations are shown as 2$\sigma$ spreads in black while the high resolution simulations are shown individually in red and green. In blue dotted and dashed lines we show the observed luminosity function for the satellites of M31 and the MW, respectively, from \citealt{2012AJ....144....4M}. Note that the observations are likely to be incomplete below $L_{\rm V} < 10^{5} L_{\odot}$.}
 \label{fig:sat_lf}
 \end{figure}

The luminosity function of satellites in the simulated LGs are shown in Figure~\ref{fig:sat_lf}. The picture is similar to the stellar mass function namely a wide spread for the simulated LGs that straddle some of the observed data. The luminosity function is  well matched to the MW's satellite luminosity function over many orders of magnitude, but shows an under prediction with respect to M31 below $L_{V}\sim10^{7}L_{\odot}$. The reader is reminded that the data is only believed to be complete above $L_{V}\gsim 10^{5}L_{\odot}$ \citep[e.g. see][]{2014MNRAS.439...73Y}.

The variance of the satellite mass and luminosity function originates in the interplay between the randomness of the small scale modes introduced, the way they couple to the larger scale constrained modes, the dwarf merger history and the inherently stochastic nature of galaxy formation at these scales (namely the susceptibility of nascent dwarf galaxies to photo-evaporation due to background ionizing radiation, strangulation, various stripping mechanisms, mergers, cold gas accretion, and feedback due to stellar winds, supernovae and possibly even AGN).

Regardless of the inherent and expected variance in comparing the satellite stellar mass and luminosity functions to each other and to the data, the simulated LGs straddle the observed data, broadly reproducing the mass and luminosity distribution of satellite dwarfs seen in the LG. 

\subsubsection{Radial concentrations}

\begin{figure}
 \includegraphics[width=20pc]{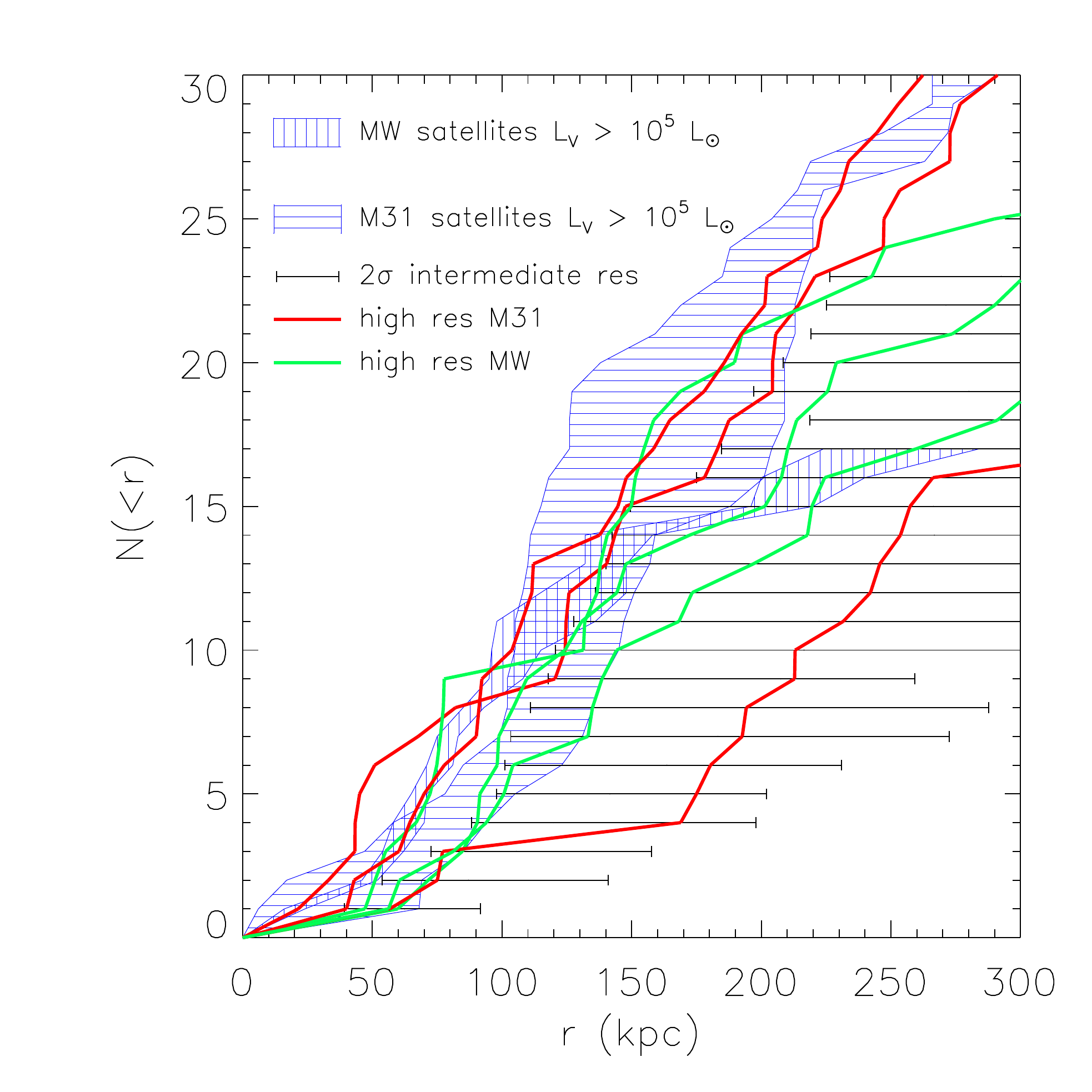}
\caption{The radial distribution of satellite galaxies whose luminosity $L_{V} > 10^{5}L_{\odot}$. This cut is chosen since it is believed to be the completeness limit for the observed LG satellites. Data for the M31 and MW are shown as the blue horizontally and vertically hatched (respectively) corridors. Observational data is taken from the compilation presented in \citet{2014MNRAS.439...73Y} updated with newly discovered dwarfs.}
 \label{fig:rad_con}
 \end{figure}

In this section and in Figure~\ref{fig:rad_con}, the radial distribution of satellite galaxies is examined and compared with the observed data. In both the simulations as well as the observations, a conservative luminosity cut is made at $L_{V}>10^{5}L_{\odot}$. Such a plot indicates how centrally concentrated or spatially extended satellite distributions around their host haloes are.  Similar to the satellite mass functions presented in the preceding section, there is a large variance for this measure. However the radial concentrations  of the two LG members \citep[data compiled by][ and updated with the dwarfs discovered since then]{2014MNRAS.439...73Y} are themselves significantly different. In the MW case at least, this may partly be explained by the difficulty in detecting satellites at low galactic latitude due to obscuration by the MW's disc (i.e. the Zone of Avoidance). On the other hand the distance to known MW satellites are known to better accuracy than for M31, hence the smaller error corridors.

Different {\sc Hestia} simulation runs have different levels of satellite concentration. In general these simulations have fewer satellites in the central regions than observed. This is, at least in part, likely a resolution effect; as the density in the inner regions of the halo increases (especially due to the presence of a dense baryonic disc), it becomes harder and harder to identify subhaloes as over densities against the background. Considering also the physical processes whereby subhaloes are destroyed \citep[e.g.][]{2020MNRAS.492.5780R}, the paucity of satellites inhabiting the inner regions, can be understood \citep[for a detailed description of this see][]{2012MNRAS.423.1200O}. The following conclusions may be drawn from Figure~\ref{fig:rad_con}: a single {\sc Hestia} run follows the M31 satellite radial distribution within the errors out to 250kpc. This in and of itself is a remarkable success. A number of other runs match the M31 data out to $\sim$200kpc. One run matches the MW data from around  40-100kpc. Most of these simulations aren't as concentrated as the observations, but the large variety in radial concentrations and the fact that  in a few cases the observations are matched, speaks to the reproducibility of this quantity. We note that \cite{2020MNRAS.491.1471S} conducted a thorough study of the radial concentrations of simulated MW mass haloes that are found both in isolation and in pairs. The results presented here are consistent with that work, as both show large variations in the radial distribution of satellites within 300 kpc of the host. It is thus unlikely that environment plays a critical role in the satellite radial distribution.

\subsubsection{The presence of a Magellanic Cloud or M33 analog}
The last near-field cosmographic feature examined and compared here is the presence of a Magellanic Cloud or M33-like object. The frequency of a Magellanic cloud like object has been studied in a number of papers \citep[see, e.g.][among others]{2011ApJ...743...40B,2011MNRAS.414.1560B}.  In fact there is a striking agreement between the frequency with which observational and theoretical studies have found Magellanic type systems around MW type galaxies. For example \cite{2011ApJ...733...62L} find that less than $\sim$10\% of isolated MW mass galaxies host such systems in the SDSS, while \cite{2010MNRAS.406..896B} find the same frequency in large cosmological simulations. Pre-Gaia estimates of the Magellanic cloud mass are around $10^{10}M_{\odot}$ \citep{2014ApJ...781..121V} while some of the more modern estimates put it at roughly an order of magnitude higher or close to 1/10th of the MW halo mass \citep{2019MNRAS.487.2685E}.  Not as much work has been done on identifying M33 type galaxies near M31 type hosts. However detailed studies attempting to study M33 like objects in zoom simulations have been carried out by, for example, \cite{2018MNRAS.480.4455M} in a bid to understand its formation. M33 is roughly the same mass as the LMC but much further away from M31 than the LMC is from the MW (230 vs. 55 kpc).

\begin{figure}
 \includegraphics[width=20pc]{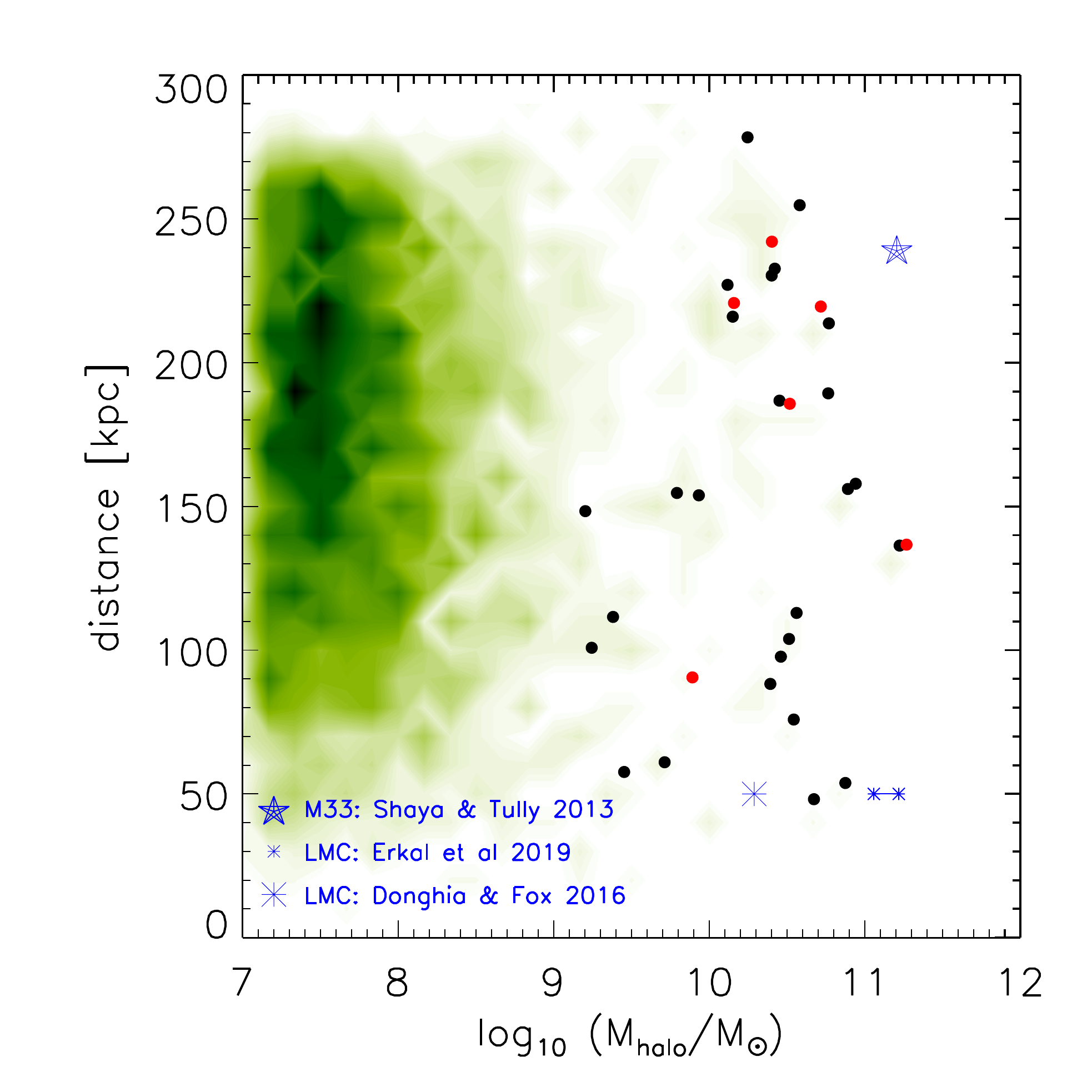}
\caption{The mass of {\sc Hestia} satellites plotted against the distance from their host center is shown here as a heat map with black regions denoting more satellites and white regions denoting no satellites. In black and red we show the most massive satellite in each the intermediate and high resolution run, respectively - namely a Magellanic cloud or M33 analog. The blue asterixes denote two observational LMC data points: the small connected ones are from employing data from \citet{2019MNRAS.487.2685E}  {\it Gaia DR2} and the large one is from the seminal review of \citet{2016ARA&A..54..363D}. The blue star is the observational data for M33 from \citet{2013MNRAS.436.2096S}.}
 \label{fig:LMC}
 \end{figure}

Figure~\ref{fig:LMC} shows the mass of satellites as a function of the distance from their host in the form of a ``heat map'', namely the darker a region is in this figure the more subhaloes have a given distance and mass.. The most massive satellite per host is shown as a circle. Some 73\% of the LG members presented here have subhaloes more massive than the lower pre-{\it Gaia} estimate, while only one ($\sim 4\%$) has an object consistent with the higher mass estimate. A few of these are at the distance of the observed Magellanic system, namely 55kpc. In fact as can be gleaned from Figure~\ref{fig:LMC}, there are  two  systems that one could liberally call ``Magellanic cloud analogs''. Similar conclusions can be made regarding an M33 analog: some massive satellites are found at distances of around 230~kpc although these are generally not as massive as the $1.6\times10^{11}{\rm M}_{\odot}$ suggested by \cite{2013MNRAS.436.2096S}.

This is not a deficit in our ability to model the Magellanic or M33 systems. The reader is reminded that structures on these scales are entirely unconstrained and thus their frequency should not {\it a priori} be any different from an unconstrained simulation of a MW mass halo. On the other hand, given that these simulations force a specific cosmographic matter distribution (namely a LG), the presence of a Magellanic cloud like object is likely to be different compared with other non-LG like environments. Thus in some cases, the LGs shown here are able to produce Magellanic Cloud or M33 like analogs.

\section{Summary and Discussion}
The properties of galaxies are known to be influenced by their environment  \citep[e.g.][]{1980ApJ...236..351D}. This is because the most basic characteristics according to which galaxies are described (such as star formation rate, merger history, color, morphology, angular momentum, baryon fraction, mass, etc.) are not only inter-dependent, but are determined, to an extent, via environmental controls \citep{2015MNRAS.448.1767C,2016MNRAS.455.2644S}. For example, galaxies without a large supply of cold gas (or where the gas supply is inhibited from cooling) will not be efficient at forming stars \citep{1991ApJ...370L..65B}. Galaxies in dense environments are likely to experience more active merger histories \citep{2012ApJ...754...26J} and thus are less likely to be rotationally supported discs. On the other hand, the current paradigm of structure and galaxy formation (namely the $\Lambda$CDM model combined with what is known theoretically and empirically about how galaxies form) has succeeded immensely when describing the large-scale properties of the galaxy distribution. Statistical properties such as $n$-point correlations functions, galaxy bimodality, or the density-morphology relationship have been understood. No existential challenges to the $\Lambda$CDM paradigm remain on large scales or when scrutinising the statistical properties of the galaxy population.

Yet challenges have been found on small scales, specifically on scales of dwarf galaxies ($\sim10^{9}M_{\odot}$) and below \citep{2017ARA&A..55..343B}. Some authors have even referred to these challenges collectively as the ``small-scale crisis of $\Lambda$CDM''. Yet what is often overlooked is that in this regime - where the smallest dwarf galaxies are examined, the only reliable data is local data, within our patch of the local Universe. The reason for this is simple: small faint, low surface brightness, low mass galaxies are only observable in our vicinity. Given that galaxy environment is known to play a role in galaxy formation, it follows that simulations that attempt to explain challenges in this regime ought to examine the ``correct'' environment.

This cogent methodology is not lost on the community at large. Most studies that attempt to explain small scale phenomena do so by examining dwarfs in MW mass haloes. Some studies, such as the ELVIS \citep{2014MNRAS.438.2578G} or the APOSTLE \citep{2016MNRAS.457.1931S,2016MNRAS.457..844F} simulations, go further and model the binarity of the LG, namely the existence of a second MW mass halo (i.e. the Andromeda system) around a~Mpc away. None of these studies explicitly control for the environment on quasi linear or linear scales - namely the existence of the Virgo cluster, the local void and the local filament, cosmographic features that individualize our local environment and regulate critical aspects of galaxy formation: merger history \citep{2019MNRAS.tmp.2688C}, halo growth \citep{2016MNRAS.460.2015S}, gas supply and bulk motion \citep{2015MNRAS.449.4494H}. Indeed, a comprehensive approach describing how the local cosmic web regulates galaxy formation, was put forward by \cite{2019OJAp....2E...7A} who examined in general detail how the cosmic web imprints itself on galaxy properties.

The issue at stake here is of a Copernican nature. To what extent the LG is fair representative of objects of its kind, and thereby to what extent its observed properties represent LG-like objects at large. This where constrained simulations of the LG are making the difference - by controlling both the actual local environment and by experimenting with the physics of galaxy formation, on the computer.

It is within this context that we have embarked on the {\sc Hestia} project - High resolution Environmental Simulations of The Immediate Area. Accordingly, the CosmicFlows-2 catalog \citep{2013AJ....146...86T} of around 8~000 peculiar velocities is employed to characterise the local cosmography.  ICs are then generated which are constrained to reproduce the $z=0$ large-scale structure, namely the gravitational attractors and repellers that constitute voids and cluster in the local Universe. Non-linear phenomena - such as the existence of a galaxy pair at the correct location with respect to the local cosmography - are not constrained but are reproduced through trial and error at low resolution but are more likely to be found in constrained simulations than in random simulations \citep{2015MNRAS.449.4494H}. The ICs are run with the cosmologiocal MHD code {\sc arepo} \citep{2010MNRAS.401..791S} and the Auriga galaxy formation model of \cite{2017MNRAS.467..179G}.

The result of the procedure is a set of 13 intermediate and 3 high resolution cosmological magneto-hydrodynamical simulation of the LG within the dynamical environment as described by measurements of the peculiar velocity field. The Virgo cluster is reproduced in the correct location and at the correct distance as is a void region analogous to the Local Void.  A filamentary structure, abutted by the simulated Local Void and stretched from the LG to the Virgo cluster is also reproduced. Within this environment, pairs of galaxies are found at roughly the correct distance from each other and from Virgo, with roughly the correct mass and mass ratio, negative relative velocity, and with cold gas discs. The simulated rotation curves of the LG analogs are in rough agreement with the observations. Dwarf galaxies, as well as galactic disc features (such as bars, spiral arms, etc) are resolved. The satellite galaxies have similar stellar mass distributions and radial concentration as observed. The {\sc Hestia} runs also produce large Magellanic cloud and M33 like objects.

The present HESTIA simulations successfully reproduce many of the observed features of the local universe on scales ranging from the $10^{14} M_{\odot}$ Virgo cluster more than 10 Mpc away, the local filament (within which the LG resides) on the scale of a few Mpc, the inter-LG sub-Mpc scale and  all the way down to  sub-virial scales within the MW and M31 galaxies. The large scales are strongly constrained by the CF2 velocity data and are robustly reproduced by the simulations. Going down to smaller and smaller scales, the constraining power of the data diminishes and the variance within the different realizations increases. The conclusion from this work is that the wide range of LG properties reported here - including the inner structure of the MW and M31, stellar and light distribution, distribution of satellites, environment - is reproduced within the constrained variance. Some of these properties are reproduced by some of the galaxies - not necessarily all of them. This opens the door to further studies on the nature of the LG and its formation. Three immediate routes are to be pursued: the increase of the ensemble of constrained simulations (thereby studying the statistical significance of our results); the increase of resolution (thereby studying the numerical robustness of the results); and experiments with the model of galaxy formation physics. These will enable a better understanding the physics of galaxy formation and to address the  Copernican issue regarding our role in the universe. The present work will be extended along these routes.  

\section*{Acknowledgments}
NIL acknowledges financial support of the Project IDEXLYON at the University of Lyon under the Investments for the Future Program (ANR-16-IDEX-0005). NIL \& PW acknowledge support from the joint Sino-German DFG research Project ``The Cosmic Web and its impact on galaxy formation and alignment'' (DFG-LI 2015/5-1). NIL, MSP, and PW thank the DAAD for PPP grant 57512596 funded by the BMBF. CP acknowledges support by the European Research Council under ERC-CoG grant CRAGSMAN- 646955. YH has been partially supported by the Israel Science Foundation grant ISF 1358/18. HC acknowledges support from the Institut Universitaire de France and the CNES. AK and GY are supported by MICIU /FEDER (Spain)  under research grant  PGC2018-094975-C21. AK further acknowledges support from the Spanish Red Consolider MultiDark FPA2017-90566- tREDC and thanks Wire for 154. ET was supported by ETAg grant IUT40-2 and by EU through the ERDF CoE TK133 and MOBTP86. SP is supported by the program ``LPI new scientific groups'' N01-2018. MV acknowledges support through an MIT RSC award, a Kavli Research Investment Fund, NASA ATP grant NNX17AG29G, and NSF grants AST-1814053, AST-1814259 and AST-1909831.
The authors gratefully acknowledge the Gauss Centre for Supercomputing e.V. (\url{www.gauss-centre.eu}) for funding this project by providing computing time on the GCS Supercomputer SuperMUC at Leibniz Supercomputing Centre (\url{www.lrz.de}). The authors also acknowledge Joss Bland-Hawthorn and Ortwin Gerhard for their help in quantifying the structural parameters of the Galaxy. Azadeh Fattahi is also acknowledged for her help in providing some of the observational LG data.

\bibliography{main}

\bsp	% typesetting comment
\label{lastpage}
\end{document}